\documentclass{article}

\usepackage{arxiv}

\usepackage[utf8]{inputenc} 
\usepackage[T1]{fontenc}    
\usepackage{hyperref}       
\usepackage{url}            
\usepackage{booktabs}       
\usepackage{amsfonts}       
\usepackage{nicefrac}       
\usepackage{microtype}      
\usepackage{lipsum}		
\usepackage{graphicx}
\usepackage[super]{natbib}
\usepackage{doi}
\usepackage{threeparttablex}
\usepackage{sistyle}
\SIthousandsep{,}

\usepackage{import}

\def\bZ{{\mathbf Z}}
\def\bbeta{{\boldsymbol{\beta}}}

\usepackage{amsthm}

\usepackage{booktabs}
\usepackage{amsmath}

\newcommand{\betahat}{\hat{\beta}}
\newcommand{\bbetahat}{\hat{\pmb{\beta}}}

\newcommand{\Vhat}{\widehat{\textrm{V}}}

\usepackage{mathtools}
 
\newcommand{\bU}{\pmb{U}}

\newcommand{\bX}{\pmb{X}}  

\newcommand{\0}{\mathbf{0}}
\newcommand{\bI}{\pmb{I}}

\newcommand{\V}{\textrm{V}}
\newcommand{\balpha}{\boldsymbol\alpha}

\usepackage{dsfont}

\title{Two-phase validation sampling via principal components to improve efficiency in multi-model estimation from error-prone biomedical databases}

\author{ \href{https://orcid.org/0000-0001-5380-2427}{\includegraphics[scale=0.06]{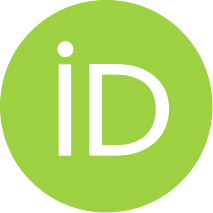}\hspace{1mm}Sarah C.~Lotspeich} \\
	Department of Statistical Sciences, Wake Forest University, Winston-Salem, North Carolina, U.S.A. \\
	\texttt{lotspes@wfu.edu} \\
	\And
    Cole ~Manschot \\ 
    Biostatistics and Research Decision Sciences, Merck \& Co., Inc., Rahway, New Jersey, U.S.A. \\
}

\renewcommand{\shorttitle}{Two-phase validation sampling via principal components}

\hypersetup{
pdftitle={ETS on PCA},
pdfsubject={q-bio.NC, q-bio.QM},
pdfauthor={Sarah C.~Lotspeich, Cole ~Manschot},
pdfkeywords={chart review, extreme tail sampling, measurement error, partial validation, principal component analysis, two-phase design},
}

\begin{document}
\maketitle

\begin{abstract}
Two-phase sampling offers a cost-effective way to validate error-prone covariate measurements in biomedical databases. Inexpensive or easy-to-obtain information is collected for the entire study in Phase I. Then, a subset of patients undergoes cost-intensive validation (e.g., expert chart review) to collect more accurate data in Phase II. When balancing primary and secondary analyses, competing models and priorities can result in poorly defined objectives for the most informative Phase II sampling criterion. Extreme tail sampling (ETS), wherein patients with the smallest and largest values of a particular quantity (like a covariate or residual) are selected, can offer great statistical efficiency in two-phase studies when focusing on a single analytic objective by targeting observations with the biggest contributions to the Fisher information. We propose an intuitive, easy-to-use approach that extends ETS to balance and prioritize explaining the largest amount of variability across multiple models of interest. Using principal components analysis, we succinctly summarize the inherent variability of all models' error-prone exposures. Then, we sample patients with the most extreme values of the first principal component for validation. Through extensive simulations and an application to the National Health and Nutrition Examination Survey (NHANES), the proposed strategy offered simultaneous efficiency gains across multiple models of interest. Its advantages persisted across various real-world scenarios, including correlated or heterogeneous measurement error. When designing a validation study, concentrating on a single model may be short-sighted. Strategically allocating resources more broadly balances multiple analytical goals simultaneously. Employing dimension reduction before sampling will allow this  strategy to scale up well to big-data applications with many error-prone exposures.
\end{abstract}



\section{Introduction}
\label{s:intro}

Biomedical data are rarely collected for a single purpose in research. Observational databases, like electronic health records (EHR), compile troves of information on extensive samples, providing new opportunities to explore various novel scientific interests with a single data source. Randomized experiments, such as clinical trials, attempt to simultaneously address the primary endpoint alongside multiple secondary objectives. Surveillance datasets, like those derived from population-based surveys, seek to describe many aspects of public health from a consolidated data collection effort. Moreover, data from 
these different study types require validation for their own reasons and in their own ways. 

\subsection{Necessity of validation in different types of studies}

Observational data are generally routinely collected with other purposes (e.g., providing good clinical care) at the forefront.\cite{Nordo2019} They may have errors resulting from various sources, like improper collection, transcription, or data cleaning.  Thus, when repurposed for research, these data may exhibit measurement error \cite{Goldstein2025} and bias downstream analyses.\cite{giganti2020} Manual chart reviews, wherein clinical experts compare research data to clinical source documents (like patients' electronic charts), can validate observational datasets,\cite{duda2012, Lotspeich2023, lotspeich2025overcomingdatachallengesenriched, shepherd2022} 
but they are usually implausible for entire studies. Partial validation (i.e., of a subset of patients) is more realistic. 

Source document verification (data validation) has long been standard in clinical trials,\cite{weiss1998} making data collected therein of generally high quality in terms of typos and other human errors. 
Further, under random assignment, the exposure of interest (generally a treatment) should be perfectly collected. Still, other variables, like biomarkers, may be imperfectly measured. Validation in the form of re-measurement using more accurate instruments can account for errors,\cite{Han2020, lyles2011, tang2015} but the more accurate measures are usually invasive or expensive, making them unreasonable to obtain for all trial participants. 

Surveillance studies can share some of the same ``human error'' elements that affect observational studies. For example, survey respondents may respond incorrectly due to recall bias. However, the gold standard (true response) is generally not obtainable through validation for these issues. More often, surveillance studies can benefit from re-measurement, as in clinical trials, like comparing self-reported dietary intake to a more resource-intensive weighed food record (as motivates our data application). 

\subsection{Informative validation sampling strategies}

Partial validation studies are a special type of two-phase design.\cite{white1982two} In Phase I, inexpensive or readily available information is collected for all patients in the study. In Phase II, a subset of patients is chosen for validation, such that more accurate measurements are collected. Sampling for Phase II can leverage any information available for all patients in Phase I, opening the door to many targeted strategies. Often, the Phase II subset is enriched for a particular model (e.g., to minimize the variance of a model coefficient of interest). This enrichment depends on the model type and estimation procedure. Essentially, identifing which patients would be most informative to validate depends on what we plan to analyze about them and how. 

For design-based estimators, like inverse probability weighting (IPW) and mean score, multiple informative sampling designs exist to minimize the variances of a particular model coefficient.\cite{reillypepe1995, amorim2021two, mcisaac2014response} Model-based estimators, like multiple imputation and maximum likelihood estimation (MLE), allow for more design flexibility than design-based ones, since some patients can be excluded from consideration (i.e., given zero validation sampling probability). While the variances of model-based estimators do not lend themselves to intuitive, analytically-defined optimal designs (as with design-based ones), many have been proposed.\cite{amorim2021two, lotspeich2023optimal, mcisaac2014response, tao2020optimal} 

When no covariate information is available pre-validation, sampling patients with extreme outcome values (called outcome dependent or extreme tail sampling [ETS]) is typically more efficient than simple random sampling (SRS).\cite{alf1975use, allison1997transmission, feldt1961use,  lin2013quantitative, risch1995extreme, sauer2021, Zhou2007} When covariate information \textit{is} available, other ETS strategies, including sampling based on (weighted) residuals \cite{lotspeich2025overcomingdatachallengesenriched, tao2020optimal,DiGravio2024, MaoCook2023} and influence functions,\cite{amorim2021two} can offer sizable efficiency gains over SRS, as well. 

\subsection{Balancing multiple models and objectives}

Most informative validation sampling designs prioritize estimating a single coefficient well (e.g., with great precision by reducing variability). Yet, post-hoc and exploratory analyses can result in competing objectives when designing a validation study of a biomedical database. Balancing multiple analytical goals across various models leads to an ill-defined optimization problem, which complicates the optimal approach for validation sampling. Of course, a ``silver bullet'' would maximize statistical performance for all objectives (unrealistic in practice), whereas maximizing for just one comes at the detriment of the others and leads to a sub-optimal system solution. For instance, focusing on maximizing efficiency for a single model may cost us efficiency in all others. Various approaches exist to the multi-objective problem,\cite{deb2016multi} but, to our knowledge, none have been applied for two-phase designs. 

Optimally designing a validation study to efficiently estimate multiple coefficients in the \textit{same} model poses a related, yet unique challenge. This setting, wherein a single model is of interest and high statistical precision for many or all coefficients is desired, has received more attention in the two-phase literature than ours with multiple models. For example, Sauer et al. (2021) demonstrate the trade-off between focusing a two-phase design strategy on one versus many coefficients in a particular model.\cite{sauer2021} Considering optimal allocation strategies that minimize (i) the variance of a single coefficient and (ii) the trace of the coefficients' variance-covariance matrix (i.e., the sum of their variances), they show that while the latter offers smaller efficiency gains for any particular coefficient it offers better precision for the model as a whole.

Intuitively, we expect the same trade-off when estimating coefficients across multiple models: We can be great for one coefficient and poor for the rest, or we can  be good for all of them. While extending existing optimal designs from one model to multiple is possible (e.g., through the adoption of alternative optimality criteria), many depend upon functions of the moments of the coefficients of interest (often unknowable in practice). We propose a simpler approach based on the already-popular ETS design that scales up well with the number of models and variables, making it particularly appealing for big-data applications. 

Principal component analysis (PCA), a popular nonparametric dimension reduction technique, offers a promising way to adapt existing informative sampling designs to accommodate multiple modeling objectives. The principal components succinctly summarize the variability within the data. 
Herein, we explore how PCA can capture information across multiple error-prone variables, each of interest in its own statistical model, and guide targeted sampling strategies for partial validation. 

\subsection{National Health and Nutrition Examination Survey (NHANES)}

In addition to extensive simulation studies, we consider the accuracy of food recall logs and dietary measurements in the National Health and Nutrition Examination Survey (NHANES).\cite{nhanes_cdc} For the 2021--2023 cohort, two dietary recalls were conducted via phone interviews. Individuals reported the foods and drinks they had consumed within the past $24$ hours. These data could be incredibly valuable in understanding how health outcomes, like vitamin levels and vitals, that are associated with intake of particular nutrients, like calcium and saturated fat (see Table~\ref{tab:nhanes models} for examples). 

\begin{table}[t]
    \centering
    \resizebox{\columnwidth}{!}{
    \begin{tabular}{p{4cm}lp{6cm}}
        \toprule
         \textbf{Outcome} & \textbf{Dietary Intake Exposure} & \textbf{Clinical Relevance} \\
         \midrule
         $Y_1:$ Vitamin D (nmol/L) & $X_1:$ Calcium (mg) & \textit{Vitamin D facilitates calcium absorption \cite{CHRISTAKOS201125}.} \\
         $Y_2:$ Resting heart rate (bpm) & $X_2:$ Caffeine (mg) & \textit{Caffeine can affect heart rate and cardiovascular function \cite{TURNBULL2017165}.}  \\
         $Y_3:$ High-density lipoproteins (HDL) cholesterol (mg/dL) & $X_3$: Total saturated fat (g) & \textit{Saturated fat intake increases HDL cholesterol \cite{Brinton1990}.} \\
         $Y_4:$ Insulin ($\mu$U/mL) & $X_4$: Alcohol (g) & \textit{Alcohol consumption may influence insulin sensitivity \cite{Schrieks2015}.} \\
         $Y_5:$ Red blood cell folate (ng/mL) & $X_5:$ Food Folate ($\mu$g) & \textit{Low folate levels can lead to higher odds of anemia \cite{Morris2007}.} \\
         \bottomrule
    \end{tabular}}
    \caption{
    The outcome of interest, associated dietary nutrient value, and the associated source data for the five models from the National Health and Nutrition Examination Survey (NHANES) data. All exposures were sourced from the dietary interview portion of the NHANES called ``What We Eat in America.'' Abbreviations for units: nanomoles per liter (nmol/L), milligrams (mg), beats per minute (bpm), milligrams per deciliter (mg/dL), grams (g), microunits per milliliter ($\mu$U/mL), nanograms per milliliter (ng/mL), and micrograms ($\mu$g).}
    \label{tab:nhanes models}
\end{table}

However, while these food recall logs are convenient and standard practice 
for dietary intake, they are subject to recall bias, under- or over-reporting, and fallibility of portion size estimation. (See Thompson and Subar (2017) and references therein.\cite{THOMPSON20175}) Converting food intake into nutrients can be further complicated by product differences in nutrient density when individuals oversimplify food items (e.g., the amount of protein in ``yogurt''). Dietary intake from food recall logs could be re-measured with more objective means (e.g., weighed food records\cite{willett2012nutritional}), but this process is much more demanding on study participants than the food diaries. A large-scale study such as the NHANES cannot implement weighed food records due to the size, cost, and burden. Still, obtaining weighed food records for a subset of NHANES participants would provide valuable insights into how error-prone the self-reported dietary intake was and allow us to correct for these errors in our analyses. 

In our data application, we illustrate how we could estimate the relationship between error-prone dietary intake from food logs and relevant health outcomes using NHANES, comparing statistical precision under various validation sampling strategies. Since conducting validation in this study was not possible, we rely on the error-prone exposures' covariance structure from the real data (the biggest driver of PCA and our proposed sampling strategy) and simulate only the true underlying dietary intake (which would be obtainable only through the appropriate labs). 

\subsection{Overview}

We propose a new informative validation sampling strategy that prioritizes multiple modeling objectives simultaneously. Sampling depends on the first principal component of all models' error-prone exposures, and the outcome variables for the multiple models may be shared or unique. Our design is simple to implement and balances competing analytical objectives in an intuitive and easy-to-understand approach. Building upon the well-established strategy of extreme tail sampling, we target patients with the largest and smallest values of the first principal component for validation. The resulting partially validated data are analyzed using multiple imputation, but other model-based methods, like maximum likelihood estimation, could be used and should see similar benefits. In comprehensive simulations, we show that the proposed design outperforms both SRS and ETS focused on a single model in terms of statistical efficiency (i.e., the inverse of the estimated coefficients' variances) in individual models and in total across all models. The proposed design's advantages persisted across many realistic scenarios, including different exposure covariance structures, increasingly severe measurement errors, complex measurement errors types (correlated or heterogeneous), and they were evident in all validation study sizes considered (suggesting that the strategy could be beneficial for studies with a wide range of budgets). Finally, we evaluate the design's performance in an application to error-prone, self-reported food intake data from the NHANES study, where it provided lower total variability (across all models) and the narrowest confidence intervals in each individual model, relative to the same comparators.

In Section~\ref{s:method}, we outline our proposed validation study design and chosen multiple imputation estimator for the analysis. In Sections~\ref{sec:sims} and \ref{sec:nhanes}, we present results from a robust collection of simulations showing the performance of our sampling strategy and an application  to error-prone, self-reported food intake data from the NHANES study, respectively. In Section~\ref{s:discuss}, we conclude with a discussion on the practicality, strengths, and limitations of our approach.

\section{Methods} \label{s:method}

\subsection{Model and data}

Suppose we are conducting a study of $N$ patients on whom $J$ separate statistical models are of interest. For each model $j$, we have a continuous outcome $Y_{ji}$, an exposure of interest $X_{ji}$, and a vector of other confounders $\bZ_{ji}$ ($j \in \{1, \dots, J\}$, $i \in \{1, \ldots, N\}$). These variables are assumed to be related through the usual normal linear regression model: 
\begin{equation} \label{eq:mean_model}
    Y_{ji} = \beta_{0j} + \beta_{1j} X_{ji} + \bbeta_{2j} \bZ_{ji} + \epsilon_{ji},
\end{equation} 
where the random errors $\epsilon_{ji}$ are independent and identically distributed (i.i.d.) following a normal distribution with mean $=0$ and variance $=\sigma_j^2$. We are most interested in estimating $\beta_{1j}$ from each model, but all the exposures $\bX_i = (X_{1i},\dots,X_{Ji})^\top$ are difficult to measure accurately.

Instead, we observe error-prone versions  $\bX^*_{i} = \bX_{i} + \bU_{i}$ for all $N$ patients, where $\bU_{i}=(U_{1i},\dots,U_{Ji})^\top$ is the vector of additive measurement errors for patient $i$'s exposures from all $J$ models. We are planning a partial validation study, such that a subset of $n$ patients ($n < N$) will have $\bX_{i}$ measured. To satisfy the missing at random (MAR) assumption\cite{Little1992} for our ultimate analyses, validation sampling can depend only on information that is fully observed (i.e., the outcomes, error-prone exposures, and confounders). Without loss of generality, we will assume that if patient $i$ is sampled for validation, then $\bX_i$ is fully observed. We want to use these partially validated data to estimate the $J$ separate models, all of which have equal priority in this study. 

Throughout, we aim to strategically design the validation study to promote the efficiency of the coefficient estimates $\hat{\bbeta}_{1} = (\hat{\beta}_{11},\dots,\hat{\beta}_{1J})^\top$, accompanying the exposures $\bX_i$ from all $J$ models. We focus on estimating these coefficients via multiple imputation here for ease of implementation. However, other model-based estimation approaches, like MLE, should benefit from similar efficiency gains under the proposed validation study design. 

\subsection{Principal components analysis of the error-prone exposures}
\label{methods:pca}

Our proposed validation sampling strategy prioritizes patients based on the first principal component of their error-prone exposures $\bX_i^*$, denoted by $PC_{1i}^*$. This dimension reduction step simplifies the sampling criterion, as we only need to consider one value per patient ($PC_{1i}^*$) rather than all $J$ of their original exposures ($\bX_i^*$). To carry out the proposed design, the first principal component $PC_{1i}^*$ ($i \in \{1, \dots, N\}$) for all patients is first obtained using existing software, briefly outlined as follows. 

We standardize the error-prone exposures prior to calculating the principal components, letting $X_{ji}^{s*} = (X_{ji}^* - \bar{X}_{j}^*)/\hat{S}^2_{X_j^*}$, $i \in \{1, \dots, N\}$), where $\bar{X}_{j}^*$ and $\hat{S}^2_{X_j^*}$ are the sample mean and sample variance, respectively, of error-prone exposure $X_j^*$ ($j \in \{1, \dots, J\}$). This standardization step ensures that $\bar{X}_{j}^{s*} = 0$ and $\hat{S}^2_{X_j^{s*}} = 1$, which accommodates error-prone exposures on different scales or with different variability.
Then, let $\bX^{s*}$ denote the $N \times J$ matrix with rows for each patient $i$ and columns for each \textit{standardized} exposure $j$, from which the $J \times J$ covariance matrix between the error-prone exposures is calculated as $\pmb{\Sigma}_{\bX^{s*}} = (N-1)^{-1}\bX^{s*\top}\bX^{s*}$. Due to the standardization, the $J \times J$ correlation matrix between them is equal to $\pmb{\rho}_{\bX^{s*}} = \pmb{\Sigma}_{\bX^{s*}}$, as well. 

This correlation matrix the starting place for our PCA of the error-prone exposures. The eigen decomposition of $\pmb{\rho}_{\bX^{s*}}$ is given by $\pmb{\rho}_{\bX^{s*}} = \pmb{V} \pmb{\Lambda} \pmb{V}^\top$, where $\pmb{\Lambda}$ is the $J \times J$ matrix with the eigenvalues on the diagonal and $\pmb{V} = \begin{bmatrix} \pmb{v}_1 & \dots & \pmb{v}_J\end{bmatrix}$ is the $J \times J$ matrix of $J \times 1$ eigenvectors $\pmb{v}_j$. Then, a patient's first principal component of their $\bX^*_i$ is obtained from this decomposition as 
$PC_{1i}^* = \bX_i^{s*} \pmb{v}_1$, which can be used to decide how informative they would be for validation (with respect to all $J$ error-prone exposures). 

In the statistical programming language \textit{R},\cite{R} $PC_1^*$ can be readily calculating using various open-source PCA implementations. We use  including the \texttt{princomp} function in the \textit{stats} package \cite{R} to conduct the PCA on the correlation matrix, as described. 

The impact of measurement error on PCA has been discussed previously. We briefly include the key findings from Hellton and Thoresen\cite{HelltonThoresen2014} here, focusing on how they relate to our proposed sampling strategy. Relative to the PCA for $\bX$, the loadings based on error-prone $\bX^*$ (i.e., coefficients in $\pmb{V}$) will have higher variability, making it potentially more difficult to interpret the true exposures' relative importance, and can be biased. We do not use the loadings directly; instead, we focus on $PC_{1i}^*$. The scores of this form, $\bX_i^{s*} \pmb{v}_j$ ($j \in \{1, \dots, J\}$), will also incur more variability when subject to measurement error. However, when using top scores, like $PC_1^*$, in analyses, the ramifications are often ``negligible,'' because the added variability from the measurement error should be small compared to the eigenvalues. Therefore, Hellton and Thoresen\cite{HelltonThoresen2014} concluded that the relative positions of the scores based on the error-prone variables closely resembled those based on the error-free. This finding is key for our validation sampling design, since it suggests that ranking patients based on $PC_1^*$ should closely approximate ranking them based on $PC_1$ (the first principal component of the true exposures $\bX$). An unattainable ETS design based on the extremes of the first principal component of the error-free exposures $PC_1$ would be ideal, in that we could try to validate patients who are most extreme with respect to all true models, and this result suggests that our proposal to use $PC_1^*$ instead is a promising alternative. 

\subsection{Validation sampling designs}\label{methods:designs}

Traditionally, targeted sampling focuses on one model $j$, and the goal is to identify which $n$ patients, if validated, would minimize the variance of $\hat{\beta}_{1j}$ corresponding to the exposure of interest $X_{ji}$.\cite{lotspeich2025overcomingdatachallengesenriched, shepherd2022, Han2020, amorim2021two, lotspeich2023optimal, mcisaac2014response, tao2020optimal, zhou2002semiparametric} Designs have also been proposed to improve estimation across multiple coefficients ($\hat{\beta}_{1j}, \bbetahat_{2j}$) in a given model $j$ by minimizing the trace (A-optimality) or determinant (D-optimality) of their variance-covariance matrix.\cite{Erdal2025, sauer2021} However, these strategies do not consider the estimation of the coefficients from the other $J-1$ models. 

Most of these designs are already complex for a single coefficient from one chosen model; extending them to consider multiple coefficients simultaneously, potentially each with its own outcome, would be challenging. Further, theoretically optimal designs to estimate $\hat{\beta}_{1j}$ are functions of true $\beta_{1j}$. Multi-wave strategies to approximate theoretically optimal designs are becoming commonplace,\cite{lotspeich2025overcomingdatachallengesenriched, shepherd2022, lotspeich2023optimal, mcisaac2015adaptive} but they can be more logistically challenging. Therefore, we instead propose a novel strategy for partial validation in a multi-objective study based on the ETS design (already a common design for single-objective studies) and using PCA to create a single measure summarizing variability across all models' error-prone exposures. 

In exploratory analyses or feature creation, PCA is a common approach to dimension reduction. Through a smaller subset of newly created variables (called principal components or loadings), PCA seeks to explain variability in the original data. This property is central to our use of PCA in validation study design. We apply PCA to all the models' error-prone exposures $\bX^*$ (and, if desired, additional confounders) to create a single variable capturing the variability between them (Section~\ref{methods:pca}). In particular, we take the first principal component $PC_1^*$, which will always explain the most variability out of all the estimated principal components. Then, we prioritize patients for validation based on $PC_1^*$, rather than focusing only on a single $X_j^*$ or trying to balance all of $\bX^*$ simultaneously. This strategy balances the reduction in variance across all $J$ models without adding computational burden, and it seamlessly integrates with existing single-objective validation sampling methods, like ETS. 

Throughout this paper, we consider three sampling strategies for the validation of $n$ patients out of a study of size $N$. As the baseline design, we include \textit{simple random sampling (SRS)}. The $n$ patients are chosen with equal probability ($n/N$), which ignores all available information and makes $\bX$ missing completely at random (MCAR).\cite{Little1992} To represent a typical existing strategy, wherein the validation study is enriched for the model of primary interest, we include \textit{extreme tail sampling on the primary model's error-prone exposure (ETS-$X_p^*$}). The original $N$ patients are ordered ascendingly by $X_p^*$ ($p \in \{1, \dots, J\}$), and those with the $n / 2$ largest and $n / 2$ smallest values are chosen for validation. In practice, the primary model would be chosen a priori, for example, based on clinical interest. Finally, we implement our proposed strategy, \textit{extreme tail sampling on the first principal component of all error-prone exposures (ETS-$PC_1^*$)}. The original $N$ patients are ordered ascendingly by $PC_1^*$, the first principal component derived from $\bX^*$ for all $J$ models, and those with the $n / 2$ largest and $n / 2$ smallest values are chosen for validation. All designs are implemented in the \texttt{auditDesignR} R package, available at \url{https://github.com/sarahlotspeich/auditDesignR}.

\subsection{Why target the extreme tails}\label{methods:intuition}

The motivation for ETS designs comes from the belief that extreme observations (i.e., those in the tails of the distribution) are most informative to conducting statistically efficient analyses. The statistical underpinnings behind this intuition stem from the large-sample theory behind MLE (the method by which common statistical models are fit). Since multiple imputation (as implemented herein) is similarly a parametric, model-based approach, the following rationale should still hold.

\subsubsection{For the primary model}

Recall that the primary model of interest captures the relationship between $Y_p$ and $X_p$ given $\bZ_p$. In linear regression, observations with exposure values $X_p$ that are farthest from the mean $\overline{X}_p$ make the largest contributions to the Fisher information for the corresponding model coefficient $\beta_{1p}$ and help to reduce variability in its estimate $\betahat_{1p}$. Thus, these extreme values of $X_p$ are deemed ``most informative'' to model estimation, and, if we can only measure the exposure for some subset of $n$ patients, focusing on those the smallest and largest values should offer the best statistical precision. (This fourth implausible strategy could be denoted by ETS-$X_p$.) 

\begin{figure}
    \centering
    \includegraphics[width=0.9\linewidth]{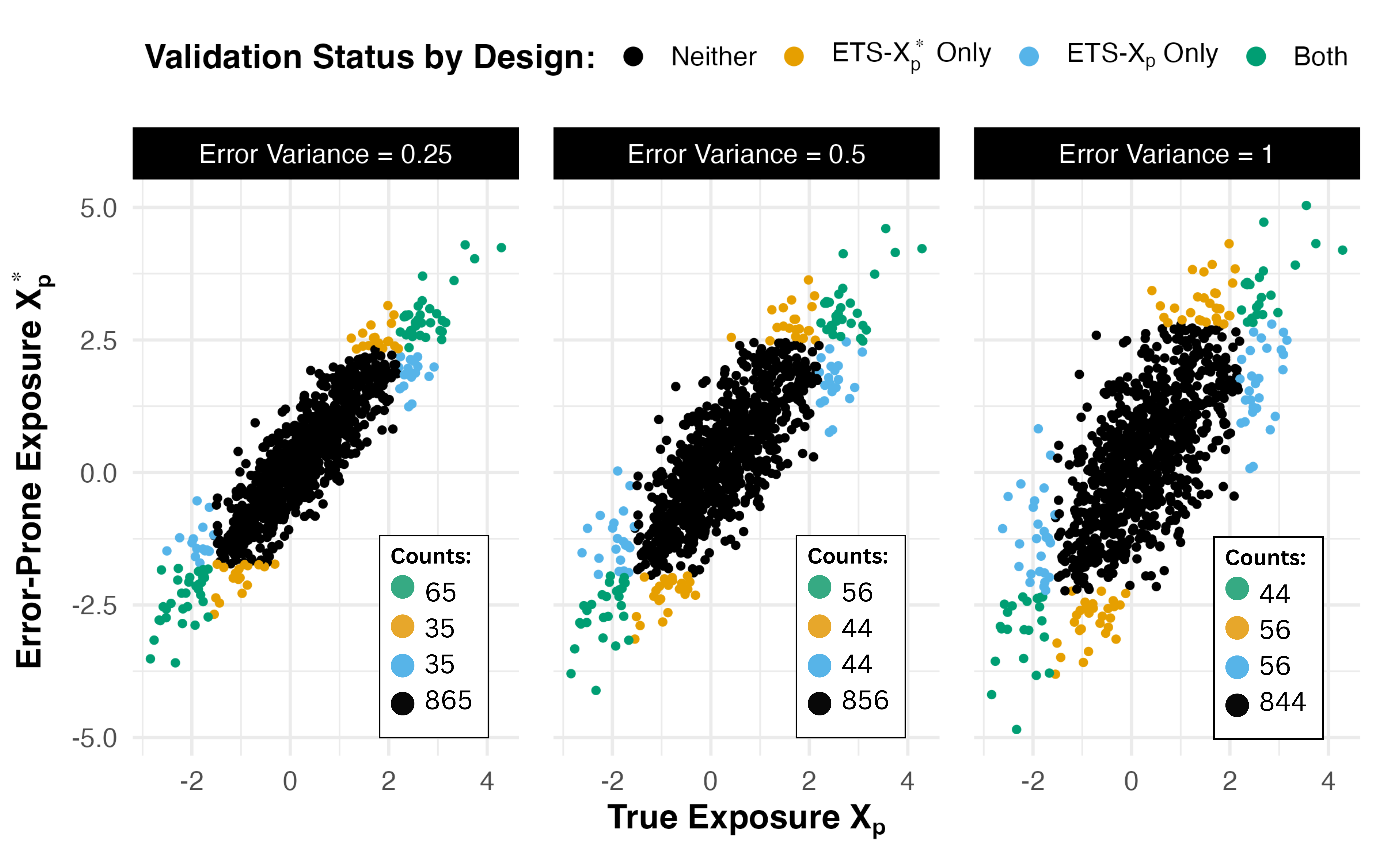}
    \caption{Plot of the primary model's true exposure $X_p$ against the error-prone version $X_p^* = X_p + U$ under varied error severities (driven by the variance of the errors $U$). Each plot contains $N = 1000$ simulated observations, and the points are colored by their validation status based on extreme tail sampling on $X_p^*$ (ETS-$X_p^*$) versus extreme tail sampling on $X_p$ (ETS-$X_p$). ``Counts'' denotes the number of observations in each setting with the corresponding validation status. As the error variance increases, and $X_p^*$ becomes less informative about $X_p$, the number of observations that would be sampled by both ETS designs decreases, and the efficiency of ETS-$X_p^*$ is expected to drop below that of ETS-$X_p$. Still, even under the worst error setting, the ETS-$X_p^*$ design could capture more information than simple random sampling (SRS).}
    \label{fig:etsXvsetsXstar}
\end{figure}

However, we cannot prioritize patients for validation based on their values of $X_p$ (which is attainable only through validation), and the ETS-$X_p^*$ design, based on the error-prone version, is perhaps the most natural alternative. The hope is that by sampling on the extremes of $X_p^*$, we can capture many of the same patients as we would have under the  ETS-$X_p$ design. The efficiency advantages of the ETS-$X_p^*$ relative to the ETS-$X_p$ will depend on how informative $X_p^*$ is about $X_p$. See Figure~\ref{fig:etsXvsetsXstar} for an illustration, wherein we compare which patients would be selected for validation based on sampling from the extremes of the true versus error-prone exposure under increasing error severities. As $X_p^*$ becomes less informative about $X_p$ under worsening measurement error, the number of observations that would be sampled by both ETS designs decreases, and the efficiency of ETS-$X_p^*$ is expected to drop below that of ETS-$X_p$. Still, even under the worst error setting, the ETS-$X_p^*$ design could capture more information than SRS. 

\subsubsection{For multiple models}

With competing analytic objectives, we shift our design focus to reducing the sum of the variances for coefficients $\betahat_{1j}$ $(j\in\{1,\ldots, J\})$ across all models. Hereafter, this sum is referred to as the total coefficient variability across all models. Improving the estimation of $\beta_{1p}$ in the primary model, through the ETS-$X_p^*$ design or otherwise, does not guarantee any improvement in the estimation of coefficients for any other models, and thus it will not help to minimize this total. As before, the most informative observations to each individual model would be those with the most extreme exposures $\bX$. Still, these values are not available when designing the validation stage, so the next-best thing is the error-prone versions $\bX^*$. However, there is now an added challenge of tailoring the design strategy to prioritize the most extreme values across multiple variables at once.

\begin{figure}
    \centering
    \includegraphics[width=0.9\linewidth]{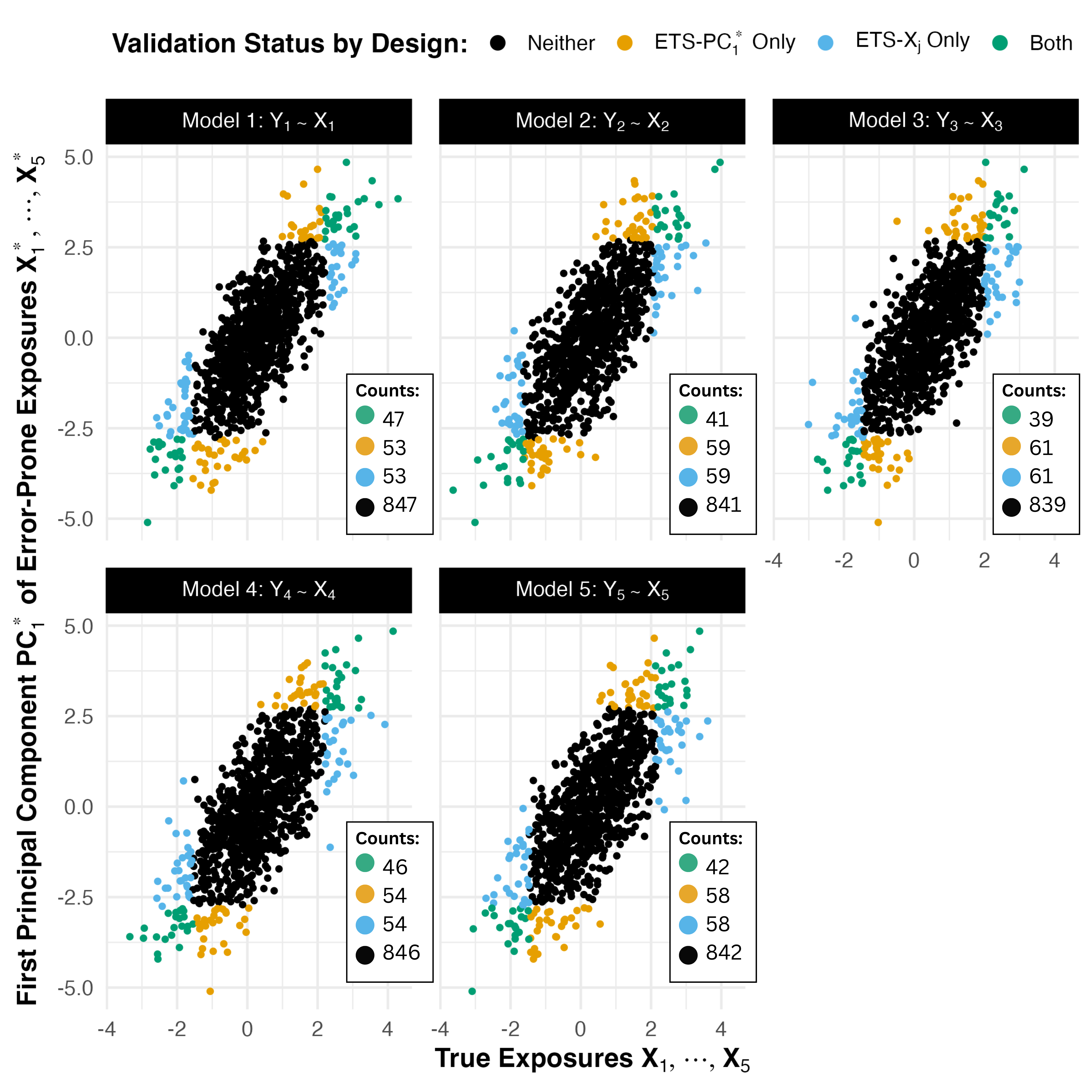}
    \caption{Plot of all models' true exposure $X_j$ against the first principal component $PC_1^*$ summarizing all the error-prone exposures $\bX^*$, focusing on the moderate error setting (error variance $= 0.5$). Each plot contains the same $N = 1000$ simulated observations, and the points are colored by their validation status based on extreme tail sampling on the first principal component $PC_1^*$ (ETS-$PC_1^*$) versus extreme tail sampling on the specific model's true exposure $X_j$ (ETS-$X_j$). ``Counts'' denotes the number of observations in each setting with the corresponding validation status. Across all five models, the number of overlapping observations that would be sampled by both ETS designs is relatively stable, and the efficiency of ETS-$PC_1^*$ is expected to offer similarly good approximations to that of the ETS-$X_j$ for each model separately.}
    \label{fig:etsXvsetsPCstar}
\end{figure}

To this end, we consider how to describe the variability across across all $\bX^*$ in a single dimension and then incorporate that information into validation sampling. Conducting PCA and taking the first principal component, $PC_1^*$, we obtain a single transformed variable that describes the greatest variability in the set of exposures $\bX^*$, providing a compact criterion on which to sample while balancing all modeling objectives. Our ETS-$PC_1^*$ design is similarly motivated as the ETS-$X_p^*$ design. By expressing the entirety of $\bX^*$ in a lower dimension as $PC_1^*$, we seek to validate patients with the most extreme values of $\bX^*$ (and, ideally, $\bX$), thereby improving estimation efficiency across all $J$ models. Essentially, we want our ETS-$PC_1^*$ to overlap, in terms of which patients are chosen for validation, with as many of the model-specific ETS-$X_j$ designs as possible. In Figure~\ref{fig:etsXvsetsPCstar}, we broaden the simulated example from above to evaluate five models of interest, focusing on the moderate error setting (error variance $= 0.5$). 
Across all five models, the number of overlapping observations that would be sampled by both ETS designs is relatively stable (between $39-47\%$ of the validation study). With this overlap, the efficiency of ETS-$PC_1^*$ is expected to offer similarly good approximations to the ETS-$X_j$ for each model separately.

The ETS-$PC_1^*$ design's ability to reduce the total coefficient variability across all models depends on two characteristics of the error-prone exposure data. First, as in the primary model case, the efficiency gains are driven by how much information we can glean from the subset of validated exposures $\bX$. When the errors in $\bX^*$ are less noisy, sampling on its extremes is expected to help us capture more informative observations in the extremes of $\bX$. Second, our ability to sample the most extreme observations with respect to all of $\bX^*$ (and subsequently $\bX$) depends on how much variability is explained by the first principal component $PC_1^*$. When the error-prone exposures $\bX^*$ are highly correlated, $PC_1^*$ will capture more variability and sampling on its extremes will lead to higher efficiency gains. In the conservative case where the $\bX^*$ are completed uncorrelated, the PCA will yield a separate factor for each exposure, and the ETS-$PC_1^*$ design will be equivalent to an ETS-$X_j^*$. In practice, when the exposures are completely independent, nominal correlation between them due to random chance can result in a slight inefficiency when using the proposed ETS-$PC_1^*$ design compared to sampling on a single variable as with ETS-$X_j^*$.

\subsection{Imputation procedure}\label{methods:imputation}

In our validated subset, we can estimate the $j$th linear regression model in \eqref{eq:mean_model} using the observed exposure $X_{ji}$ directly. For the rest of the patients in the study, we replace their missing value with an informative placeholder $\widehat{X}_{ji}$. Various imputation procedures for measurement error settings like this one, where partial validation data are available, have been proposed.\cite{giganti2020, Cole2006, Edwards2013, Edwards2015, Edwards2020, Shepherd2012, Han2021, Pelgrims2023, Amorim2024} Specifically, we adopt a single deterministic imputation framework in the Simulations (Section~\ref{sec:sims}). (Single imputation is computationally quicker than multiple, and we focused on the estimates' empirical variability to compare the designs.) However, multiple imputation is preferred if estimated variances are desired (e.g., for inference), as in the NHANES application (Section~\ref{sec:nhanes}), in which case the imputation models need to be modified to ensure congeniality with the outcome models. R code for imputation can be found online at \url{https://github.com/sarahlotspeich/ETS_PCA}.

\subsubsection{Different imputation models under different validation study designs}\label{subsec:imp_mods}

For an unvalidated patient, $\widehat{X}_{ji}$  comes from an imputation model fit to the validation study. Any fully available information that helps estimate $X_{ji}$ can be incorporated, and any variables used in the validation study design \textit{must} be included.\cite{zhou2002semiparametric} Under an SRS design, the imputed values are defined simply from the error-prone exposure $X_{ji}^*$ and the additional confounders $\bZ_{i}$ as 
\begin{align}
\widehat{X}_{ji} = \widehat{\alpha}_{0j} + \widehat{\alpha}_{1j} X^*_{ji} + \widehat{\balpha}_{2j} \bZ_i. \label{imp_vals:srs}
\end{align}
When $X_{ji}$ is numeric, the estimated coefficients in \eqref{imp_vals:srs} can be obtained via normal linear regression. Of course, different regression models could be used to impute other types of exposures. 

If the validation study was selected under an ETS-$X_p^*$ design, focused on the primary model, the error-prone exposure $X_p^*$ used in the design must be included in \textit{all} exposures' imputation models ($p \in \{1, \dots, J\}$). Then, the imputed values $\widehat{X}_{ji}$ are: 
\begin{align}
\widehat{X}_{ji} &= 
\begin{cases}
\widehat{\alpha}_{0j} + \widehat{\alpha}_{1j} X^*_{ji} + \widehat{\balpha}_{2j} Z_i \text{ if } j = p \text{, and } \\ 
\widehat{\alpha}_{0j} + \widehat{\alpha}_{1j} X^*_{ji} + \widehat{\balpha}_{2j} Z_i + \widehat{\alpha}_{3j} X_{pi}^* \text{ otherwise.} 
\end{cases} \label{imp_vals:etsX*}
\end{align}
Note that the coefficients will take on different values in the first versus second case. Under our proposed ETS-$PC_1^*$ design, the first principal component $PC_1^*$ must be included in all imputation models. The imputed values $\widehat{X}_{ji}$ are then defined as 
\begin{align}
\widehat{X}_{ji} &= \widehat{\alpha}_{0j} + \widehat{\alpha}_{1j} X^*_{ji} + \widehat{\alpha}_{2j} Z_i + \widehat{\alpha}_{3j} PC^*_{1i}. \label{imp_vals:etsPC1*}
\end{align}
As with the SRS design, estimating the imputation models under either ETS design depends on the type of exposures, but can be done using standard regression techniques. For ease of notation, $\widehat{\balpha}$ are the estimated coefficients for any of the imputation models. 

Single imputation following \eqref{imp_vals:srs}--\eqref{imp_vals:etsPC1*} has been called \textit{deterministic imputation}, since missing values are replaced with a single model prediction $\widehat{X}_{ji}$ and treated as fixed in the subsequent analyses.\cite{DAgostinoMcGowan2024} This approach can be implemented using existing software to (i) fit the imputation model and (ii) calculate the predictions. Alternatively, the \texttt{mice} function in the \textit{mice} R package can be used with argument \texttt{method = "norm.predict"}.\cite{mice} 

\subsubsection{Modifications for multiple imputation} \label{subsec:mi_vs_si}

When using multiple imputation, the imputation models need to be modified one step further. Due to the stochastic nature of multiple imputation, which incorporates variability about the imputation procedure itself, the analysis model outcomes $Y_{ji}$  must be included in the imputation models. Otherwise, the imputation models will not be \textit{congenial} with the analysis model, and the resulting multiple imputation estimates will be biased.\cite{DAgostinoMcGowan2024,Moons2006} 

There are various ways one might incorporate randomness into the multiple rounds of imputation. 
One of the earliest and most common methods is to (i) resample the coefficients $\widetilde{\balpha}$ from a distribution based on the fitted ones $\widehat{\balpha}$ and their variance-covariance matrix $\widehat{\Sigma}$ in the validation study and (ii) generate the imputed values $\widehat{X}_{ji}$ from the distribution dictated by $\widetilde{\balpha}$.\cite{rubin2004multiple} For example, if $X_{ji}$ is continuous, then in each imputation the parameters $\widetilde{\balpha}$ will first be randomly drawn from a multivariate Normal distribution with mean vector $\widehat{\balpha}$ and variance-covariance matrix $\widehat{\Sigma}$. This process is repeated many times, the analysis model is fit to each imputed dataset, and the estimates from each fitted model are ultimately pooled using Rubin's rules. We can again use the \texttt{mice} R package. Specifically, the \texttt{mice} function can be used with argument \texttt{method = "norm"}, which will include the outcome $Y_{ij}$ in the imputation model by default. 

\section{Simulations}\label{sec:sims}

\subsection{Data generating mechanism}\label{sec:dgm}

We generated data for samples of $N = 1000$ patients, $n$ of whom had their exposures validated ($n \in \{100, 150, 250\}$).  Each patient had a binary covariate $Z$ generated from a Bernoulli distribution with $\Pr(Z=1)=0.3$. Then, true exposures $\pmb{X}=(X_1, X_2, X_3, X_4, X_5)$ were generated from a multivariate normal distribution with mean vector $=\0_5$ and one of three covariance matrices: \textit{independence:} $\pmb{\Sigma}_X = \bI_5$, 
\begin{align*}
    \textit{equal dependence:} \quad \pmb{\Sigma}_X = \begin{bmatrix} 
    1.00 & 0.50 & 0.50 & 0.50 & 0.50 \\
    0.50 & 1.00 & 0.50 & 0.50 & 0.50 \\
    0.50 & 0.50 & 1.00 & 0.50 & 0.50 \\
    0.50 & 0.50 & 0.50 & 1.00 & 0.50 \\
    0.50 & 0.50 & 0.50 & 0.50 & 1.00
    \end{bmatrix}, &\textit{ or } \\
    \textit{unequal dependence:} \quad \pmb{\Sigma}_X = \begin{bmatrix} 
    1.00 & 0.05 & 0.10 & 0.20 & 0.35 \\
    0.05 & 1.00 & 0.15 & 0.25 & 0.40 \\
    0.10 & 0.15 & 1.00 & 0.30 & 0.45 \\
    0.20 & 0.25 & 0.30 & 1.00 & 0.50 \\
    0.35 & 0.40 & 0.45 & 0.50 & 1.00
    \end{bmatrix},
\end{align*}
where $\0_5$ was a $5 \times 1$ vector of zeros and $\bI_5$ the $5 \times 5$ identity matrix. 
Each exposure in $\pmb{X}$ was related to its own numeric response:
\begin{align}
    \pmb{Y} &= \begin{bmatrix} Y_1 \\ Y_2 \\ Y_3 \\ Y_4 \\ Y_5 \end{bmatrix} = \begin{bmatrix} 0.5X_1+0.1Z \\ 1+X_2+0.2Z \\ 2+1.5X_3+0.3Z \\ 3+2X_4+0.4Z \\ 4+2.5X_5+0.5Z \end{bmatrix} + \pmb{\epsilon}
\end{align}
for a vector of independent random errors $\pmb{\epsilon}$ generated from a standard multivariate normal distribution. The error-prone versions of these exposures were constructed as $\pmb{X}^* = \pmb{X} + \pmb{U}$, where the measurement errors $\pmb{U}$ were generated from a multivariate normal distribution with mean vector $=\0_5$ and covariance matrix $= \sigma_U^2 \pmb{I}_5$ ($\sigma_U^2 \in \{0.25, 0.5, 1\}$). In additional simulations, correlated measurement errors and ones where the variance $\sigma_U^2$ differed by variable were also considered (Sections~S1.1--S1.2 in the Supplementary Materials). 

\subsection{Metrics for comparison}

To select the $n$ simulated patients for the validation study, the SRS, ETS-$X_1^*$, and ETS-$PC_1^*$ designs were considered. Without loss of generality, the model of $Y_{1}$ given $(X_1,Z)$ was treated as \textit{primary}. When discussing individual models, we focus on comparing the empirical efficiency for the coefficient estimates $\widehat{\bbeta}_{1}$ corresponding to the exposures $\pmb{X}$. When evaluating design performance across all modeling objectives simultaneously, we examine the sum of these coefficients' empirical variances. 

For each of $1000$ simulated datasets per setting, separate linear regression models were estimated following the single imputation approach from Section~\ref{methods:imputation}. These replicates are summarized by the empirical efficiencies for $\widehat{\beta}_{1j}$ ($j \in \{1, 2, 3, 4, 5\}$) in each of the five models (i.e., the inverse of the empirical variances, $\Vhat(\betahat_{1j})^{-1}$) and the empirical total coefficient variability across all models, (i.e., the sum of all models' empirical variances, $\sum_{j=1}^{5}\Vhat(\betahat_{1j})$). The imputation estimator was empirically unbiased under all settings, models, and validation study designs. To guide discussion, we focus on graphical comparisons of the different designs' efficiencies and total coefficient variability in the main text. Full numerical details (including bias, empirical standard errors, and empirical relative efficiency to SRS [RE]) can be found in the Supplementary Materials.

\subsection{Exposures' covariance structure}\label{sims:vary_cov_struct}

We explored three different covariance structures (independence, equal dependence, and unequal dependence) for the exposures $\pmb{X}$ described in Section~\ref{sec:dgm} and fixed the error variance $\sigma^2_U = 1$ and validation study size $n = 100$. Since PCA is expected to be informative for correlated variables, the ETS-$PC_1^*$ design is expected to perform best when the exposures $\bX^*$ are dependent. 

\begin{figure}
    \centering
    \includegraphics[width=0.9\linewidth]{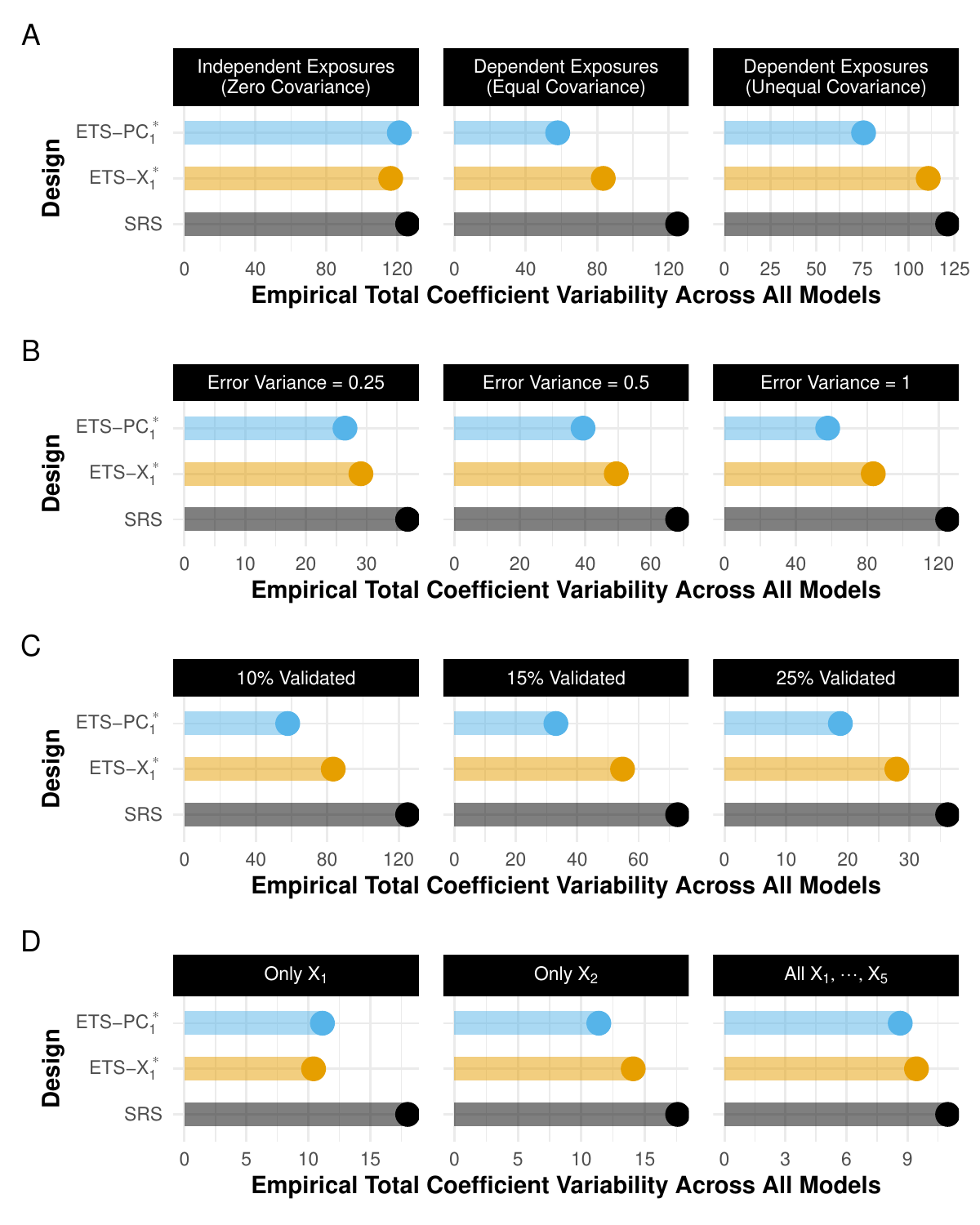}
    \caption{Simulation results comparing the empirical total coefficient variability across all models $\sum_{j=1}^{5}\widehat{\V}(\hat{\beta}_{1j})$ under simple random sampling (SRS), extreme tail sampling on the primary error-prone exposure $X_1^*$ (ETS-$X_1^*$), and extreme tail sampling on the first principal component of all the error-prone exposures $X_1^*,\dots,X_5^*$ (ETS-$PC_1^*$) validation study designs. In A, three different covariance structures for the five exposures $X_1, \dots, X_5$ were considered. In B, three different variances $\sigma_U^2$ for the additive measurement errors (assumed to be uncorrelated and homogeneous) in exposures $X_1^*, \dots, X_5^*$ were considered. In C, three different proportions of validated patients out of $N = 1000$ were considered. In D, there was a shared outcome $Y$, and it was generated from the exposures $X_1, \dots, X_5$ under scenarios where only one exposure was associated (only $\beta_{11} \neq 0$ for $X_1$ or only $\beta_{12} \neq 0$ for $X_2$) versus all exposures are associated (all $\beta_{1j} \neq 0$ for $X_1,\dots,X_5$).}
    \label{fig:sumvar_barbell}
\end{figure}

The imputation estimator was empirically unbiased ($\leq 1\%$) for all models, designs, and covariance structures (Supplemental Table~S1). As expected, the total empirical variability across all models was smallest for the ETS-$PC_1^*$ design when the exposures $\bX$ were dependent (either equally or unequally), such that $PC_1^*$ was informative about $\bX^*$ and, to a lesser extent, $\bX$ (Figure~\ref{fig:sumvar_barbell}A). Under equal dependence, the total coefficient variability for the ETS-$PC_1^*$ design was only $58$, compared to $83$ for the ETS-$X_1^*$ design and $125$ for SRS. When assuming unequal dependence instead, the total coefficient variability for all designs was larger, but their relative advantages persisted: $76$ for ETS-$PC_1^*$ versus $111$ for ETS-$X_1^*$ and $121$ for SRS. Only when the exposures were independent did the ETS-$X_1^*$ design lead to a slightly smaller total ($116$ versus $121$), and ETS-$PC_1^*$ still beat SRS (at $126$).

\begin{figure}
    \centering
    \includegraphics[width=0.9\linewidth]{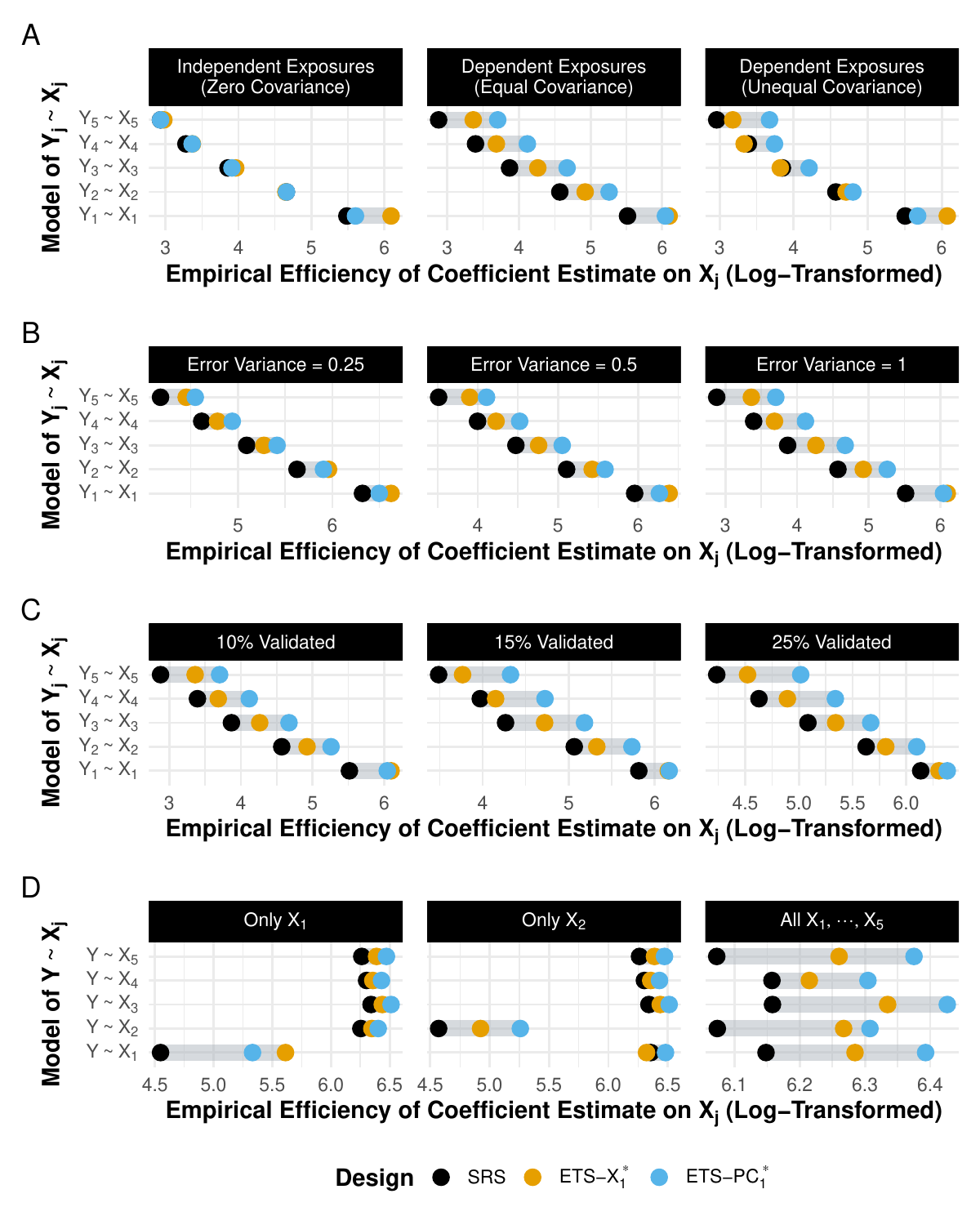}
    \caption{Simulation results comparing the per-model empirical efficiency under simple random sampling (SRS), extreme tail sampling on the primary error-prone exposure $X_1^*$ (ETS-$X_1^*$), and extreme tail sampling on the first principal component of all the error-prone exposures $X_1^*,\dots,X_5^*$ (ETS-$PC_1^*$) validation study designs. In A, three different covariance structures for the five exposures $X_1, \dots, X_5$ were considered. In B, three different variances $\sigma_U^2$ for the additive measurement errors in exposures $X_1^*, \dots, X_5^*$ (assumed to be uncorrelated and homogeneous) were considered. In C, three different proportions  of validated patients out of $N = 1000$ were considered. In D, there was a shared outcome $Y$, and it was generated from the exposures $X_1, \dots, X_5$ under scenarios where only one exposure was associated (only $\beta_{11} \neq 0$ for $X_1$ or only $\beta_{12} \neq 0$ for $X_2$) versus all exposures are associated (all $\beta_{1j} \neq 0$ for $X_1,\dots,X_5$).}
    \label{fig:all_barbell}
\end{figure}

When broken down by model, there were clear efficiency gains with the ETS-$PC_1^*$ design for all secondary models (of $Y_2, \dots, Y_5$) under equal or unequal dependence (Figure~\ref{fig:all_barbell}A), with RE of $1.26-2.28$ versus $1.15-1.48$ for the ETS-$X_1^*$ design. These gains for the ETS-$PC_1^*$ design were comparable in size across all models under equal dependence, and there was hardly any efficiency loss to the ETS-$X_1^*$ design for the primary model. Under unequal dependence, the largest efficiency gain was seen for the model of $Y_5$, which had the exposure $X_5^*$ with the highest correlations, on average, to the others. Under unequal covariance, the efficiency for the primary analysis when sampling with ETS-$PC_1^*$ fell farther behind the ETS-$X_1^*$ design than in other settings. However, this model-specific loss was offset by the lower total empirical variability across all models mentioned above, as desired. 
When the exposures were generated independently, ETS-$PC_1^*$ was much less efficient than ETS-$X_1^*$ for the primary model (RE $=1.13$ versus $1.83$), as expected, but still better than SRS. For the secondary models, ETS-$PC_1^*$ and ETS-$X_1^*$ performed similarly under independence, offering modest efficiency gains over SRS (RE $\leq 1.11$). 

In summary, the ETS-$PC_1^*$ can offer substantial efficiency gains both individually and across multiple models when the exposures behind $PC_1^*$ are correlated, which can be measured in practice before designing the validation study. Moreover, even if the exposures are not correlated (or are only modestly so), the ETS-$PC_1^*$ still offered better statistical precision than SRS. 

\subsection{Exposures' error severity, correlation, and homogeneity}\label{sims:vary_error_var}

In their discussion of the impact of measurement error on PCA, Hellton and Thoresen\cite{HelltonThoresen2014} distinguish between uncorrelated versus correlated (i.e., whether errors in different variables can impact each other) and homogeneous versus heterogeneous (i.e., whether the error severity is the same for all variables) errors. Through the simulations that follow, we fully examine how our proposed PCA-based sampling strategy performed under all such settings. 

First, to isolate the contribution of the severity under uncorrelated, homogeneous exposure errors, we changed the variance $\sigma_U^2 \in \{0.25, 0.5, 1\}$ used to generate the errors $\pmb{U}$ and create error-prone exposures $\pmb{X}^* = \pmb{X} + \pmb{U}$. We fixed the validation study size at $n=100$ and used the equal dependence covariance structure for $\bX$; all other variables were generated following Section~\ref{sec:dgm}. 

Again, all imputation coefficient estimates were empirically unbiased ($\leq 0.9\%$, Supplemental Table~S2). The ETS-$PC_1^*$ offered the smallest total empirical coefficient variability across all models regardless of how severe the measurement error was (Figure~\ref{fig:sumvar_barbell}B). The efficiency gap between the ETS-$PC_1^*$ and ETS-$X_1^*$ designs widened as $\sigma_U^2$ increased: from $26$ versus $29$ under the lowest error severity to $58$ versus $83$ under the highest. Intuitively, the need for validation, in general, is greater when $\sigma^2_U$ is larger, such that $\bX^*$ is noisier about $\bX$. Meanwhile, the benefits of our proposed ETS-$PC_1^*$ design over the ETS-$X_1^*$ and SRS grew as $\sigma_U^2$ increased. In other words, including information from all error-prone exposures, rather than just the primary one, seemed to become more important as the errors worsened. Both ETS designs offered better efficiency than SRS when breaking down into individual models (RE $= 1.20-2.28$ for ETS-$PC_1^*$ versus $1.18-1.79$ for ETS-$X_1^*$). As $\sigma_U^2$ increased, the efficiency of all designs and models decreased, and the advantages of the ETS-$PC_1^*$ design became more pronounced (Figure~\ref{fig:all_barbell}B). In fact, under the most severe setting, the two ETS designs offered fairly similar efficiency gains over SRS in the primary model (RE $= 1.79$ for ETS-$X_1^*$ and $1.70$ for ETS-$PC_1^*$).

Overall, the results from additional simulations with homogeneous, correlated errors or with heterogeneous, uncorrelated errors were similar (Sections~S1.1 and S1.2, respectively, in the Supplementary Materials). The imputation estimates continued to exhibit virtually no bias ($<1\%$) everywhere (Supplemental Tables~S3 and S4). In terms of efficiency, the proposed ETS-$PC_1^*$ design demonstrated sizable gains over ETS-$X_1^*$ and SRS on both a per-model and all-model basis (Supplemental Figures~S1 and S2). In individual models, the ETS-$PC_1^*$ design led to RE as high as $2.12$ under homogeneous, correlated errors and as high as $2.48$ under heterogeneous, uncorrelated errors. Under those same scenarios, the maximum RE for ETS-$X_1^*$ was only $1.66$ and $1.94$, respectively. Across models, the empirical total coefficient variability with the ETS-$PC_1^*$ design was always smaller than with the ETS-$X_1^*$ or SRS. Even for the primary model of $Y_1$ and $X_1$, the two ETS designs had similar efficiency; elsewhere, ETS-$PC_1^*$ could be much more efficient. 

Collectively, these simulations highlight how the proposed ETS-$PC_1^*$ validation study design offers persistent improvements to per-model and all-model efficiency across complex error settings one might encounter in practice: homogeneous or heterogeneous and uncorrelated or correlated. As we varied other characteristics across simulation settings, we focused on the uncorrelated, homogeneous errors elsewhere for simplicity.

\subsection{Validation study size}\label{sims:vary_val_prop}

We varied the proportion of the $N = 1000$ patients whose exposures $\bX$ were validated between $10\%$ (used elsewhere), $15\%$, and 25\%. We kept $\sigma_U^2 = 1$ for uncorrelated, homogeneous errors and used the equal dependence covariance structure for $\bX$; all other variables were generated as described in Section~\ref{sec:dgm}. 

The estimates were empirically unbiased ($\leq 0.9\%$) with even the smallest validation study considered (Supplemental Table~S5). For all settings considered, the ETS-$PC_1^*$ design offered the smallest empirical total variability across the five models (Figure~\ref{fig:sumvar_barbell}C), decreasing from $58$ to $19$ as the validation proportion increased. Meanwhile, the total variability with the ETS-$X_1^*$ design improved from $83$ to $28$, respectively. 

The gap between the ETS-$PC_1^*$ and ETS-$X_1^*$ designs' total variability was fairly consistent in magnitude across all validation study sizes considered. As the validation proportion increased, fewer patients have missing data and need imputation, leading to better efficiency for all designs. Still, incorporating all five error-prone exposures through the ETS-$PC_1^*$ design remained beneficial over focusing on a single one. In other words, regardless of how many patients were validated, the ETS-$PC_1^*$ design gained more information across all models, offering better statistical efficiency, than the ETS-$X_1^*$ and SRS. 

When evaluating the models individually, the efficiency of all designs also increased as more patients were validated, as expected (Figure~\ref{fig:all_barbell}C). The gains with the ETS-$PC_1^*$ design were persistent and large for all validation proportions considered, with RE $= 1.28-2.50$ versus $1.19-1.79$ for ETS-$X_1^*$. Together, these per-model and all-model findings suggest that the ETS-$PC_1^*$ design could be beneficial for partial validation studies with a wide range of budgets and sizes.

\subsection{Shared outcome model}

We considered an alternative data generating mechanism where the five exposures $X_1, \dots, X_5$ shared a common numeric outcome $Y$, but separate models of $Y$ given $X_j$ and $Z$ were of interest. The shared outcome was generated as $Y = \beta_1 X_1 + \beta_2 X_2 + \beta_3 X_3 + \beta_4 X_4 + \beta_5 X_5 + 0.1Z + \epsilon$, where random errors $\epsilon$ come from a standard normal distribution. We fixed the validation study size $n = 100$, error variance $\sigma_U^2=1$ (assumed to be uncorrelated, homogeneous), and used the equal dependence covariance structure for $\bX$; all other variables were generated as in Section~\ref{sec:dgm}. Three scenarios were included: (i) only $X_1$ has a relationship with $Y$ ($\beta_{11} = 1$ and all other $\beta_{1j} = 0$), (ii) only $X_2$ has a relationship with $Y$ ($\beta_{12} = 1$ and all other $\beta_{1j} = 0$), and (iii) all $\bX$ have relationships with $Y$ (all $\beta_{11} = \cdots = \beta_{15} = 0.2$). Through these settings, we compared the validation study designs when only one exposure was associated with $Y$ (and when it was/was not the one targeted with the ETS-$X_1^*$ design) versus when all exposures were associated. 

Imputation continued to lead to unbiased estimates ($\leq 1.2\%$, Supplemental Table~S6), and an interesting new story emerged when comparing the designs' efficiency. Essentially, ETS-$PC_1^*$ offered the same (if not better) precision as ETS-$X_1^*$, both in terms of the total variability across all coefficients and efficiency in the individual models (Figure~\ref{fig:all_barbell}D). When only $X_1$ was associated with $Y$, the ETS-$PC_1^*$ was highly efficient (RE $= 2.19$) but beaten by ETS-$X_1^*$ (RE $= 2.90$) in the model using $X_1$, as expected. However, the ETS-$PC_1^*$ design offered efficiency gains over ETS-$X_1^*$ for the other models (RE $=1.16-1.23$ versus $1.06-1.13$). Altogether, in this particular scenario, the efficiency ``boost'' from tailoring the design specifically to $X_1^*$, the error-prone version of the only significant biomarker for $Y$, led to a smaller total coefficient variability with ETS-$X_1^*$ than ETS-$PC_1^*$, but the two were relatively close ($10$ versus $11$). (Across all simulations considered herein, the only other time ETS-$PC_1^*$ did not have the smallest total variability was when exposures were independent.)

When only $X_2$ was associated with $Y$, the ETS-$PC_1^*$ design offered substantially better efficiency in total ($11$ versus $14$ and $18$ with ETS-$X_1^*$ and SRS, respectively) and for all five models separately. The benefits were especially clear for the model with $X_2$, where ETS-$PC_1^*$ had RE $= 1.99$, while ETS-$X_1^*$ had RE $= 1.42$. Interestingly, in this case, the ETS-$X_1^*$ design was actually slightly less efficient than SRS for the primary model of $Y$ and $X_1$ (RE $= 0.97$), while the ETS-$PC_1^*$ still offered some improvement (RE $= 1.13$).  Finally, when all $\bX$ were associated with $Y$, the ETS-$PC_1^*$ design outperformed the ETS-$X_1^*$ and SRS designs for all models together and separately (including in the primary model with $X_1$). 

Overall, these results suggest that ETS-$PC_1^*$ was the ``safest'' design choice, often offering the same (if not better) precision as ETS-$X_1^*$, unless we are very confident about which biomarker(s) are associated with the outcome. In exploratory analyses, for example, we could be likely to see benefits with the ETS-$PC_1^*$, which casts a wider net across all exposures, than if we were to focus on a single (potentially not associated) exposure with the ETS-$X_1^*$ design. 
 
\section{Application to the Error-Prone Dietary Intake Exposures} \label{sec:nhanes}

Having established the relative advantages of our proposed ETS-$PC_1^*$ validation study design in synthetic simulations (Section~\ref{sec:sims}), we further examined its performance in more realistic data. Conducting multiple validation studies under different design strategies was not feasible in practice. Instead, we adopted error-prone exposures from a real-world dataset and generated validated versions from them. Importantly, with these data, we no longer controlled the covariance between the error-prone exposures, which heavily influences how best to select our validation subset. Comparing the statistical efficiency of $\widehat{\bbeta}_{1}$ between same-size validation studies with different design strategies remained our primary objective. 

\subsection{About the survey and data}

The NHANES includes various demographic, examination, laboratory, and questionnaire data. 
We focus on the  dietary interview portion, called the ``What We Eat in America'' (WWEIA), which 
captures individual-level daily consumption of $64$ nutrients and food components and is an incredibly valuable source for dietary intake  data.\cite{Hoy2015}

In the most recent WWEIA ($2021$--$2023$), a representative sample of individuals of all ages was asked to complete two $24$-hour dietary intake recalls approximately $1$ week apart. The food components and quantities recorded were converted into dietary nutrients using the Food and Nutrient Database for Dietary Studies, which provides caloric and nutrient information.\cite{fndds} The NHANES data set reports the cumulative intake of these nutrients for each recall. 

\subsection{Modeling diet-driven health outcomes}\label{app:models}

We considered five models of interest, each one linking an examination or laboratory outcome $Y_j$ to a corresponding dietary intake exposure $X_j$ (Table~\ref{tab:nhanes models}). Each model was adjusted by the individuals' sex, age, race and ethnicity (Hispanic origin), and education level to account for possible confounding, denoted  together by $\bZ$. Each model followed the linear regression in \eqref{eq:mean_model}, and estimating the exposure coefficients $\bbeta_1$ 
across all models was our goal.

We focused on individuals who (i) completed the first telephone interview and (ii) were not missing any of the models' outcomes, exposures, or confounders. A sample of $N = 2388$ met these criteria and were used for analysis. While excluding individuals with missing data is not always advisable (e.g., when missing data patterns may be informative), we elected to do so here to simplify our illustration of the proposed validation study design.

\subsection{Measurement error in dietary intake}\label{app:add_error}

The NHANES dietary interview uses the Automated Multiple-Pass Method (AMPM) which is considered the most sophisticated methodology for 24-hour food recall.\cite{THOMPSON20175} ADPM begins with a list of all foods and beverages consumed, followed by probing questions, collecting the time and occasion of meals,  detailed descriptions, and a final nudge for anything else consumed. 

Two main limitations of the AMPM exist. First, it requires multiple days of recalls to obtain accurate \textit{mean} measurements of intake, which is both costly and time-consuming. Second, it can induce self-report bias from psycho-social behaviors that result in under or over estimation of nutrient and energy intake, though underreporting is most common.\cite{THOMPSON20175} Measurement error may occur during the food recall and impact assessment on the relationship of dietary intake with other biomarkers. Thus, the available dietary intake data from NHANES, based on self-reported food diaries, are treated as the error-prone exposures $\bX^* = (X_1^*, X_2^*, X_3^*, X_4^*, X_5^*)$. 
Calcium ($X_1^*$) and saturated fat ($X_3^*$) were moderately correlated with each other ($\hat{\rho} = 0.59$) and weakly correlated with food folate ($X_5^*$) ($\hat{\rho} = 0.38$ and $0.34$, respectively). Caffeine ($X_2^*$) and alcohol ($X_4^*$) exhibited minimal correlation with other exposures (all $\hat{\rho} \leq 0.16$). See Supplemental Figure~S3 for the full correlation matrix. 

Weighed food records are the gold standard for recording portions and the most accurate measurements, but this method has a high burden on participants.\cite{willett2012nutritional} Weighed food records require measuring all portions prior to consumption and any food waste that remains after a meal. 
Conducting weighed food records for large studies like NHANES is not feasible. Therefore, we simulated the error-free exposures $\bX = (X_1, X_2, X_3, X_4, X_5)$, which would be obtainable only through this method, by generating  measurement errors $\pmb{U} = (U_1, U_2, U_3, U_4, U_5)$ from a multivariate normal distribution with mean vector
$= \0_5 $ and covariance matrix 
$= \mathrm{diag}(77805.2, 7630.9, 70.5, 119.1, 4991.9)$, where $\mathrm{diag}(\cdot)$ denotes a diagonal matrix. The error variance was chosen to be $\sigma_{Uj}^2 = \widehat{\textrm{Var}}(X_j^*)/4$, i.e., $1/4$ times the estimated variance of the corresponding error-prone exposure. Since each exposure had its own variance $\sigma_{Uj}^2$, this setup led to heterogeneous measurement error. Then, we added these errors to the NHANES data to get $\bX = \bX^* - \pmb{U}$, by manipulating the classical additive measurement error model to solve for $\bX$ instead of $\bX^*$. Supplemental Figure~S4 contains scatterplots of the resulting simulated $\bX$ versus the actual $\bX^*$ from NHANES. While we used uncorrelated errors here, we found in the simulations (Section~\ref{sims:vary_error_var}) that the relative performance of the different validation study designs was very similar for correlated errors. 

\subsection{Partial validation of dietary intake}\label{app:val_des}

All $N = 2388$ individuals had outcomes $\pmb{Y} = (Y_1, Y_2, Y_3, Y_4, Y_5)$, error-prone dietary intake $\pmb{X}^*$, and other confounders $\pmb{Z}$ observed (all taken from the real NHANES dataset). However, only a subset of $n = 250$ of them were treated as having error-free dietary intake $\pmb{X}$ (simulated) assumed to be obtained through validation with weighted food logs, and the particular subset of individuals chosen depended on the sampling strategy. For the remaining $N - n = 2138$ unvalidated individuals, error-free dietary intake exposures were multiply imputed $75$ times. 

First, we included SRS, which, while theoretically valid, generally results in the widest 95\% confidence intervals ($95\%$ CIs) for $\bbeta_{1}$, the coefficients on the dietary-intake exposures (of primary interest). Partial validation with SRS leads to model estimates that are asymptotically consistent (i.e., they converge in probability to the true coefficients) as both sample sizes ($N$ and $n$) go to infinity. However, there can be discrepancies between the SRS estimates and full validation (gold standard) in finite samples. 

Second, we considered an ETS-$X_1^*$ design that focuses on the error-prone dietary intake exposure, $X_1^*$ (calcium), from the first model. This approach might appeal if we were primarily interested in this model, so that we tailored our validation sampling strategy to focus on it. However, we may later decide that the other four models are also clinically relevant, and then re-purpose the validation study data to estimate them, as well. While this ETS-$X_1^*$ design could offer the narrowest $95\%$ CI for the first model, it may not even outperform SRS for the others, leading us to compromise in terms of performance across all analytic objectives. 

Third, we applied our proposed ETS-$PC_1^*$ design. As the five exposures were measured in different units and on different scales, it was critical to conduct the PCA on the correlation, as described in Section~\ref{methods:pca}, rather than on the covariance.
The first principal component, $PC_1^*$, explained $39\%$ of the variability in the error-prone dietary intake exposures $\bX^*$ from NHANES. For more detail about the variability explained by the other principal components, see the scree plot in Supplemental Figure~S5. All error-prone dietary intake exposures contributed to $PC_1^*$ (Supplemental Figure~S6), with the largest loadings for calcium ($X_1^*$), saturated fat ($X_3^*$), and food folate ($X_5^*$), which had the highest correlations. There were smaller loadings for caffeine ($X_2^*$), followed by alcohol ($X_4^*$), both of which had relatively little correlation with the other exposures. 

For comparison, we calculated the average percent variability explained by $PC_1^*$ in simulations. When varying the covariance structures for $\bX$, $28\%$ was explained under independence, $48\%$ under equal dependence, and $39\%$ under unequal dependence. As the exposure error variance $\sigma_U^2$ increased (for uncorrelated, homogeneous errors), the percent of variability explained by $PC_1^*$ dropped from $64\%$ to $48\%$. For the uncorrelated, heterogeneous errors, the percent of variability explained when $\sigma_U^2 = 1$ but different exposures were allowed to be more/less error-prone remained between $48-55\%$, on average. Based on the correlation matrix (Supplemental Figure~S3), the NHANES data resembled the unequal dependence scenario; thus, the $39\%$ variability explained by $PC_1^*$ aligned  well with the simulations in this regard. Plus, these data were slightly more challenging than the simulations in that the magnitude of heterogeneous measurement errors in the dietary intake exposures was fairly severe, and some of the exposures had very little correlation with the others. 

\subsection{Results from multiple models}

\begin{figure}
    \centering
    \includegraphics[width=\linewidth]{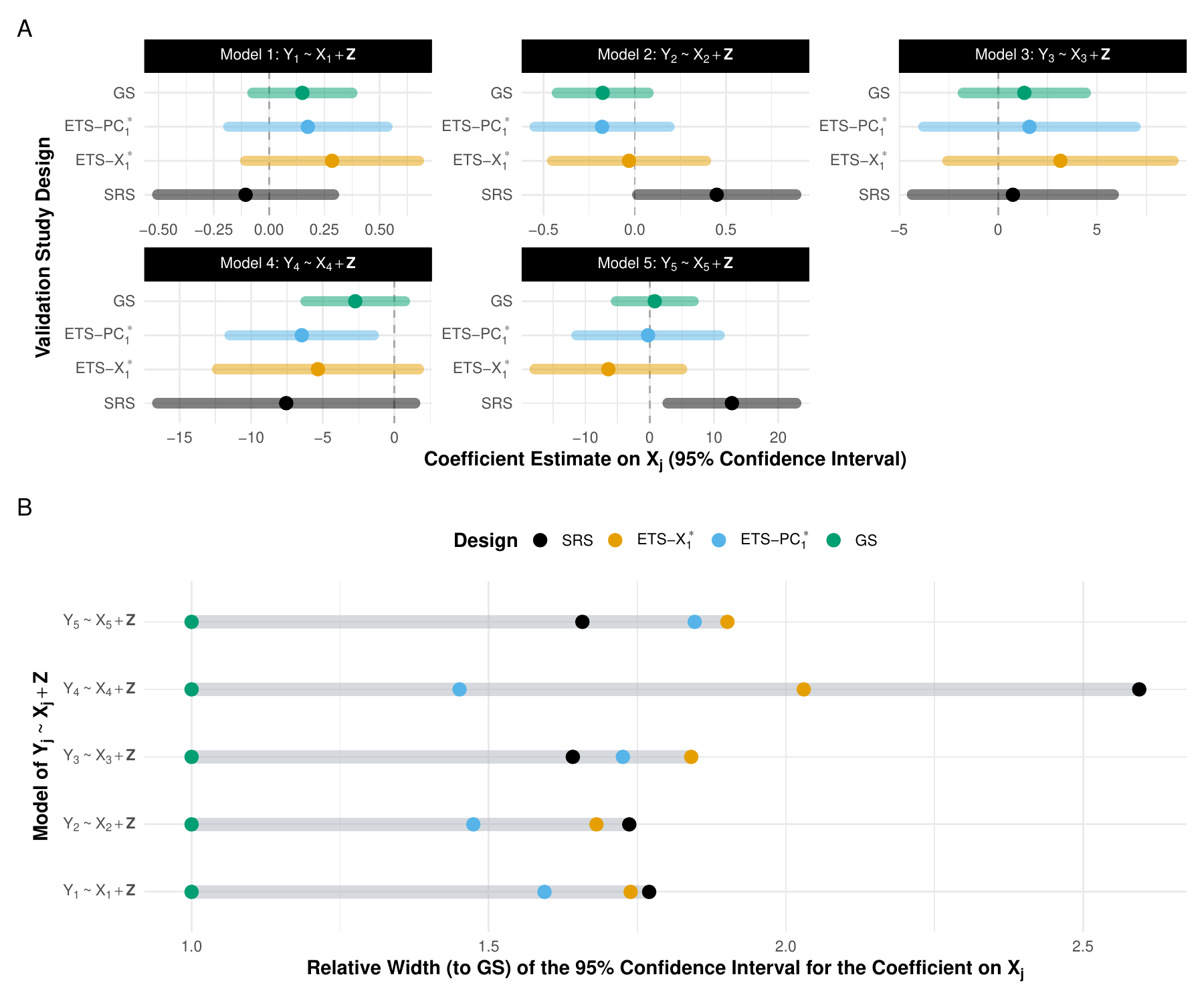}
    \caption{
    Coefficient estimates and corresponding $95\%$ confidence intervals (A) and relative widths of $95\%$ confidence intervals to the gold standard analysis (B) for $\beta_{1j}$ ($j \in \{1, \dots, 5\}$), the coefficients corresponding to each of the five dietary intake exposures from the five diet-driven health outcome models fit to the National Health and Nutrition Examination Survey (NHANES) data. We considered full validation (all $N = 2388$ individuals) to be the gold standard and compared it to partial validation for $n = 250$ of $N = 2388$, where individuals could be chosen via simple random sampling (SRS), extreme tail sampling on the primary error-prone exposure $X_1^*$ (ETS-$X_1^*$), and extreme tail sampling on the first principal component of all the error-prone exposures $X_1^*,\dots,X_5^*$ (ETS-$PC_1^*$). All partially validat3ed analyses multiply imputed missing validated exposures for the remaining individuals $75$ times.}
    \label{fig:forest}
\end{figure}

The coefficient estimates for the dietary intake exposures from all five models are displayed in Figure~\ref{fig:forest}A. For interpretability, all exposures were rescaled from their original NHANES units just before fitting the models (Table~\ref{tab:nhanes models}). Calcium, caffeine, and food folate originally had maximums in the thousands, so they were transformed to $100$-unit increments. Saturated fat and alcohol only had maximums in the hundreds, so they were instead rescaled to $10$-unit increments. In addition to the partially validated analyses, we also included the gold standard (fully validated) analysis, which used simulated $\bX$ for all individuals as a benchmark. Between different validation study designs, the magnitude of some estimated associations differed slightly. In particular, SRS could have a reversed direction to the ETS designs, and, in instances where this happened, SRS disagreed with the gold standard analysis, as well. 

The total coefficient variability for the partially validated analysis using SRS was $\sum_{j=1}^{5}\widehat{V}(\hat{\beta}_{1j}) = 53.9$, which was actually slightly lower than that of $55.7$ using the ETS-$X_1^*$ design. Since the primary model of $Y_1$ and $X_1$ seemed to capture only a slight association (even for the gold standard analysis), it makes sense that the tailoring our design to that model may not be beneficial here. With the ETS-$PC_1^*$ design, the total coefficient variability was smallest of the partially validated analyses at $46.4$. 

Additionally, the ETS-$PC_1^*$ design offered narrower $95\%$ CIs than SRS or ETS-$X_1^*$ in most models (Figure~\ref{fig:forest}B). It had the largest advantages in Model $4$, which was interesting given that $X_4^*$ (alcohol) was not very correlated with the other exposures and had the smallest loading on $PC_1^*$. However, according to the gold standard, this association seemed to be strongest of the five considered (i.e., it had the largest $\hat{\beta}_{1j}$ in magnitude), so perhaps that made the benefit easier to see. In Models $1-3$ and $5$, the ETS-$X_1^*$ design did not offer much improvement over SRS in terms of CI width and could even be worse. Interestingly, the one model where it \textit{did} seem to help was not the the one corresponding with $X_1$. While unexpected, particularly given the simulation results, the error-prone dietary intake exposure could be so noisy that sampling on the extremes of $X_1^*$ did not necessarily lead us to validate the extremes of $X_1$, which would be the real source of the efficiency gain. Further, the estimated coefficient on $X_1$ in this model seemed to be the smallest of the five exposures considered. 

The 95\% CIs for the adjusted associations between calcium, caffeine, and saturated fat and their corresponding outcomes (from the Models $1-3$) contained the null value of zero for the gold standard and all partially validated analyses. In the Model $4$, the designs all agreed that increasing alcohol was associated with lower insulin, on average, after adjusting for other factors. However, the smaller variability with the ETS-$PC_1^*$ design led to a narrower 95\% CI that was fully bounded away from zero, while the ETS-$X_1^*$ and SRS designs did not. (Due to its smaller point estimate, the gold standard interval actually contained zero here, too.) 
Finally, in Model $5$, the SRS analysis suggested that higher food folate was associated with higher folate, on average, after adjusting for other individual characteristics, but the other designs and the gold standard analysis disagreed and led to either null or negative associations. For summaries of the estimated coefficients on other variables, see Table~\ref{tab:nhanes_fitted_models}. 

\section{Conclusion} \label{s:discuss}

Extreme tail sampling on the first principal component offers an accessible, intuitive way to balance multiple, competing analytical priorities when choosing which individuals to validate in studies with exposure measurement error. This strategy amplifies the benefits of a conventional single-objective two-phase design by avoiding the expenses of conducting separate, targeted validation samples for distinct models. It does so without incurring computational complexity, and we provide the user-friendly \texttt{auditDesignR} R package at \url{https://github.com/sarahlotspeich/auditDesignR} with an easy-to-use implementation. 

In simulations, our proposed sampling strategy improved the overall efficiency of the two-phase estimates across multiple models of interest under mild assumptions. The improvement was evident under varying covariance structures; measurement error severities and complexities (correlation and heterogeneity); and validation proportions. Models where the error-prone exposure was strongly correlated with the others saw the greatest improvement. 

We demonstrate the ETS-$PC_1^*$ design's advantages in single and multiple imputation analyses, where model specification is critical to achieving unbiased estimates. In practice, selection procedures (e.g., based on information criteria) can help select the ``best-fitting'' imputation model.\cite{SCHOMAKER2014758} However, the advantages of sampling on the first principal component should also manifest if applied to other model-based estimation procedures that can handle targeted sampling, which opens up possibilities for more robust analytical approaches, like semiparametric MLE.\cite{tao2021} 

When conducting multi-wave validation studies,\cite{shepherd2022,lotspeich2023optimal,mcisaac2015adaptive} the ETS-$PC_1^*$ design could be used for initial waves. Using the preliminary information on $\bX$ and $\bX^*$ collected thus far, it could be possible to adopt an alternative method to conventional PCA (used here) that accounts for measurement error.\cite{Sanguinetti2005,Wentzell2012} Then, sampling for later validation waves could be based on the first of these ``corrected'' principal components, perhaps offering additional efficiency gains by reducing noise between what we sample on (the principal component) and what we use for analysis (the validated exposures).  

Toward broader applicability of the proposed sampling strategy, there are multiple other interesting directions for future work. First, investigations into efficiency gains with other common outcome types (including binary, count, or time-to-event) are needed. Second, adopting other existing designs for a dimension-reduced version of the error-prone exposures (like the first principal component) could offer further advantages. Third, modifications to other error settings (e.g., error-prone outcomes) are promising; in particular, the set of variables contributing to the PCA will need to be revisited. 

\bibliographystyle{wileyNJD-AMA}
\bibliography{bibliography}






\section*{Supplementary Materials}
\begin{itemize}
    \item \textbf{Additional appendices, tables, and figures:} The supplemental figures and tables referenced in Sections 3--4 are available online at \url{https://github.com/sarahlotspeich/ETS_PCA/blob/main/Supplement.pdf} as Supplementary Materials.
    \item \textbf{R-package for sampling designs:} An \textsf{R} package \texttt{auditDesignR} that implements the various validation study designs in this article is available at \url{https://github.com/sarahlotspeich/auditDesignR}.
    \item \textbf{R code and data:} The R scripts and data needed to replicate the simulation studies and NHANES analysis are available at \url{https://github.com/sarahlotspeich/ETS_PCA}.
\end{itemize}

\begin{table}
\centering
\resizebox{\columnwidth}{!}{
\begin{tabular}{lcccc}
\toprule
\textbf{Coefficient} & \textbf{GS} & \textbf{SRS} & \textbf{ETS-$\pmb{X_1^*}$} & \textbf{ETS-$\pmb{PC_1^*}$}\\
\midrule
\addlinespace
\multicolumn{5}{c}{\textit{Model $1$: Vitamin D ($Y_1$) and Calcium ($X_1$)}}\\
\addlinespace
Intercept & $43.24$ ($36.99$, $49.50$) & $46.38$ ($39.00$, $53.76$) & $41.67$ ($34.32$, $49.03$) & $43.08$ ($36.12$, $50.05$)\\
Exposure of Interest: Calcium & $0.00$ ($0.00$, $0.00$) & $0.00$ ($-0.01$, $0.00$) & $0.00$ ($0.00$, $0.01$) & $0.00$ ($0.00$, $0.01$)\\
Female (Referent = Male) & $12.33$ ($9.54$, $15.11$) & $11.78$ ($8.83$, $14.72$) & $12.59$ ($9.73$, $15.45$) & $12.27$ ($9.47$, $15.07$)\\
Age & $0.73$ ($0.64$, $0.81$) & $0.72$ ($0.63$, $0.80$) & $0.73$ ($0.65$, $0.82$) & $0.73$ ($0.64$, $0.81$)\\
\addlinespace[0.3em]
\multicolumn{5}{l}{Race and Ethnicity (Referent = Non-Hispanic White)}\\
\hspace{1em}Mexican American & $-14.59$ ($-20.40$, $-8.77$) & $-14.8$ ($-20.64$, $-8.96$) & $-14.65$ ($-20.46$, $-8.84$) & $-14.61$ ($-20.43$, $-8.79$)\\
\hspace{1em}Other Hispanic & $-10.58$ ($-15.36$, $-5.81$) & $-10.96$ ($-15.75$, $-6.16$) & $-10.28$ ($-15.09$, $-5.46$) & $-10.46$ ($-15.29$, $-5.64$)\\
\hspace{1em}Non-Hispanic Black & $-18.78$ ($-23.49$, $-14.06$) & $-19.29$ ($-24.09$, $-14.5$) & $-18.45$ ($-23.22$, $-13.67$) & $-18.86$ ($-23.58$, $-14.14$)\\
\hspace{1em}Other Race & $-6.65$ ($-11.36$, $-1.93$) & $-6.92$ ($-11.67$, $-2.18$) & $-6.61$ ($-11.33$, $-1.88$) & $-6.80$ ($-11.52$, $-2.08$)\\
\addlinespace[0.3em]
\multicolumn{5}{l}{Education Level (Referent = College Graduate or Above)}\\
\hspace{1em}$<9$th Grade & $-7.35$ ($-14.97$, $0.27$) & $-7.52$ ($-15.14$, $0.10$) & $-7.18$ ($-14.83$, $0.46$) & $-7.28$ ($-14.92$, $0.35$)\\
\hspace{1em}$9-11$th Grade & $-15.27$ ($-21.30$, $-9.24$) & $-15.39$ ($-21.45$, $-9.34$) & $-15.49$ ($-21.52$, $-9.45$) & $-15.23$ ($-21.26$, $-9.20$)\\
\hspace{1em}High School Grad/GED or Equivalent & $-7.25$ ($-11.02$, $-3.48$) & $-7.50$ ($-11.31$, $-3.70$) & $-7.28$ ($-11.05$, $-3.5$) & $-7.16$ ($-10.96$, $-3.36$)\\
\hspace{1em}Some College or AA Degree & $-2.71$ ($-6.04$, $0.61$) & $-2.65$ ($-5.98$, $0.68$) & $-2.73$ ($-6.06$, $0.60$) & $-2.64$ ($-5.97$, $0.68$)\\
\addlinespace
\multicolumn{5}{c}{\textit{Model $2$: Resting Heart Beat ($Y_2$) and Caffeine ($X_2$)}}\\
\addlinespace
Intercept & $73.61$ ($71.61$, $75.62$) & $72.55$ ($70.45$, $74.65$) & $73.34$ ($71.3$, $75.38$) & $73.58$ ($71.54$, $75.63$)\\
Exposure of Interest: Caffeine & $0.00$ ($0.00$, $0.00$) & $0.00$ ($0.00$, $0.01$) & $0.00$ ($0.00$, $0.00$) & $0.00$ ($-0.01$, $0.00$)\\
Female (Referent = Male) & $3.11$ ($2.14$, $4.08$) & $3.34$ ($2.36$, $4.32$) & $3.19$ ($2.23$, $4.16$) & $3.16$ ($2.19$, $4.13$)\\
Age & $-0.12$ ($-0.15$, $-0.09$) & $-0.12$ ($-0.15$, $-0.09$) & $-0.12$ ($-0.15$, $-0.09$) & $-0.12$ ($-0.15$, $-0.09$)\\
\addlinespace[0.3em]
\multicolumn{5}{l}{Race and Ethnicity (Referent = Non-Hispanic White)}\\
\hspace{1em}Mexican American & $-2.83$ ($-4.87$, $-0.80$) & $-2.46$ ($-4.52$, $-0.41$) & $-2.75$ ($-4.81$, $-0.68$) & $-2.83$ ($-4.87$, $-0.78$)\\
\hspace{1em}Other Hispanic & $0.41$ ($-1.26$, $2.08$) &  $0.81$ ($-0.89$, $2.5$) & $0.50$ ($-1.18$, $2.18$) & $0.45$ ($-1.22$, $2.12$)\\
\hspace{1em}Non-Hispanic Black & $-1.02$ ($-2.68$, $0.64$) & $-0.34$ ($-2.05$, $1.37$) & $-0.86$ ($-2.53$, $0.80$) & $-0.98$ ($-2.65$, $0.69$)\\
\hspace{1em}Other Race & $0.64$ ($-1.01$, $2.28$) & $0.73$ ($-0.92$, $2.38$) & $0.69$ ($-0.97$, $2.35$) & $0.62$ ($-1.03$, $2.28$)\\
\addlinespace[0.3em]
\multicolumn{5}{l}{Education Level (Referent = College Graduate or Above)}\\
\hspace{1em}$<9$th Grade & $1.28$ ($-1.38$, $3.94$) & $1.55$ ($-1.12$, $4.23$) & $1.31$ ($-1.35$, $3.97$) & $1.29$ ($-1.38$, $3.95$)\\
\hspace{1em}$9-11$th Grade & $3.24$ ($1.13$, $5.35$) & $3.16$ ($1.02$, $5.29$) & $3.24$ ($1.13$, $5.35$) & $3.24$ ($1.13$, $5.34$)\\
\hspace{1em}High School Grad/GED or Equivalent & $2.58$ ($1.26$, $3.89$) & $2.68$ ($1.36$, $4.01$) & $2.59$ ($1.27$, $3.91$) & $2.57$ ($1.26$, $3.89$)\\
\hspace{1em}Some College or AA Degree & $2.73$ ($1.57$, $3.90$) & $2.55$ ($1.38$, $3.72$) & $2.69$ ($1.5$, $3.88$) & $2.75$ ($1.57$, $3.92$)\\
\addlinespace
\multicolumn{5}{c}{\textit{Model $3$: HDL Cholesterol ($Y_3$) and Saturated Fat ($X_3$)}}\\
\addlinespace
Intercept & $46.34$ ($43.74$, $48.95$) & $46.54$ ($43.48$, $49.6$) & $45.55$ ($42.22$, $48.87$) & $46.24$ ($43.15$, $49.34$)\\
Exposure of Interest: Saturated Fat & $0.01$ ($-0.02$, $0.04$) & $0.01$ ($-0.04$, $0.06$) & $0.03$ ($-0.03$, $0.09$) & $0.02$ ($-0.04$, $0.07$)\\
Female (Referent = Male) & $8.86$ ($7.70$, $10.01$) & $8.82$ ($7.64$, $10.00$) & $9.06$ ($7.81$, $10.31$) & $8.87$ ($7.68$, $10.06$)\\
Age & $0.11$ ($0.08$, $0.15$) & $0.11$ ($0.08$, $0.15$) & $0.11$ ($0.08$, $0.15$) & $0.11$ ($0.08$, $0.15$)\\
\addlinespace[0.3em]
\multicolumn{5}{l}{Race and Ethnicity (Referent = Non-Hispanic White)}\\
\hspace{1em}Mexican American & $-3.61$ ($-6.01$, $-1.2$) & $-3.59$ ($-6.03$, $-1.14$) & $-3.52$ ($-5.94$, $-1.1$) & $-3.54$ ($-5.98$, $-1.1$)\\
\hspace{1em}Other Hispanic & $-2.00$ ($-3.99$, $-0.01$) & $-2.06$ ($-4.07$, $-0.06$) & $-1.81$ ($-3.86$, $0.24$) & $-1.95$ ($-4.0$, $0.10$)\\
\hspace{1em}Non-Hispanic Black & $2.79$ ($0.84$, $4.74$) & $2.76$ ($0.80$, $4.71$) & $2.84$ ($0.88$, $4.80$) & $2.78$ ($0.83$, $4.74$)\\
\hspace{1em}Other Race & $-0.14$ ($-2.09$, $1.82$) & $-0.18$ ($-2.13$, $1.78$) & $-0.07$ ($-2.05$, $1.9$) & $-0.16$ ($-2.11$, $1.8$)\\
\addlinespace[0.3em]
\multicolumn{5}{l}{Education Level (Referent = College Graduate or Above)}\\
\hspace{1em}$<9$th Grade & $-6.40$ ($-9.55$, $-3.25$) & $-6.42$ ($-9.58$, $-3.27$) & $-6.27$ ($-9.44$, $-3.1$) & $-6.44$ ($-9.59$, $-3.28$)\\
\hspace{1em}$9-11$th Grade & $-5.00$ ($-7.50$, $-2.51$) & $-4.98$ ($-7.48$, $-2.48$) & $-5.01$ ($-7.51$, $-2.52$) & $-4.96$ ($-7.46$, $-2.47$)\\
\hspace{1em}High School Grad/GED or Equivalent & $-3.48$ ($-5.04$, $-1.92$) & $-3.48$ ($-5.04$, $-1.92$) & $-3.40$ ($-4.97$, $-1.83$) & $-3.46$ ($-5.03$, $-1.90$)\\
\hspace{1em}Some College or AA Degree & $-3.30$ ($-4.68$, $-1.92$) & $-3.29$ ($-4.67$, $-1.91$) & $-3.37$ ($-4.76$, $-1.99$) & $-3.31$ ($-4.69$, $-1.93$)\\
\bottomrule
\end{tabular}
}
\caption{All fitted models' estimates from the application to error-prone dietary intake exposures in the National Health and Nutrition Examination Survey (NHANES) data. For the gold standard (GS) analysis, all $N = 2388$ individuals were assumed to have validated exposures $\bX$ (simulated and assumed to be obtainable only via validation). Otherwise, a subset of $n = 250$ individuals were assumed to have $\bX$; for all others, the validated exposures $\bX$ were multiply imputed from $\bX^*$ and $\bZ$. Three different validation study designs were considered: simple random sampling (SRS), extreme tail sampling on the primary error-prone exposure $X_1^*$ (ETS-$X_1^*$), and extreme tail sampling on the first principal component of all the error-prone exposures $X_1^*,\dots,X_5^*$ (ETS-$PC_1^*$).}
\label{tab:nhanes_fitted_models}
\end{table}

\renewcommand{\thetable}{2 (Continued)}
\begin{table}
\centering
\resizebox{\columnwidth}{!}{
\begin{tabular}{lcccc}
\toprule
\textbf{Coefficient} & \textbf{GS} & \textbf{SRS} & \textbf{ETS-$\pmb{X_1^*}$} & \textbf{ETS-$\pmb{PC_1^*}$}\\
\midrule
\addlinespace
\multicolumn{5}{c}{\textit{Model $4$: Insulin ($Y_4$) and Alcohol ($X_4$)}}\\
\addlinespace
Intercept & $13.79$ ($10.34$, $17.25$) & $14.16$ ($10.59$, $17.73$) & $14.15$ ($10.58$, $17.72$) & $14.24$ ($10.73$, $17.76$)\\
Exposure of Interest: Alcohol & $-0.03$ ($-0.06$, $0.01$) & $-0.08$ ($-0.17$, $0.01$) & $-0.05$ ($-0.12$, $0.02$) & $-0.06$ ($-0.12$, $-0.01$)\\
Female (Referent = Male) & $-2.57$ ($-4.26$, $-0.87$) & $-2.67$ ($-4.39$, $-0.95$) & $-2.78$ ($-4.53$, $-1.04$) & $-2.84$ ($-4.57$, $-1.11$)\\
Age & $-0.01$ ($-0.06$, $0.04$) & $0.00$ ($-0.06$, $0.05$) & $-0.01$ ($-0.06$, $0.04$) & $-0.01$ ($-0.06$, $0.04$)\\
\addlinespace[0.3em]
\multicolumn{5}{l}{\textit{Race and Ethnicity (Referent = Non-Hispanic White)}}\\
\hspace{1em}Mexican American & $3.39$ ($-0.16$, $6.95$) & $3.34$ ($-0.24$, $6.91$) & $3.28$ ($-0.3$, $6.86$) & $3.17$ ($-0.4$, $6.75$)\\
\hspace{1em}Other Hispanic & $3.19$ ($0.28$, $6.10$) & $3.04$ ($0.07$, $6.00$) & $3.2$ ($0.29$, $6.12$) & $2.99$ ($0.06$, $5.91$)\\
\hspace{1em}Non-Hispanic Black & $0.46$ ($-2.41$, $3.33$) & $0.74$ ($-2.17$, $3.66$) & $0.54$ ($-2.34$, $3.42$) & $0.49$ ($-2.39$, $3.36$)\\
\hspace{1em}Other Race & $-0.91$ ($-3.79$, $1.97$) & $-0.85$ ($-3.77$, $2.06$) & $-0.84$ ($-3.72$, $2.05$) & $-0.82$ ($-3.71$, $2.07$)\\
\addlinespace[0.3em]
\multicolumn{5}{l}{Education Level (Referent = College Graduate or Above)} \\
\hspace{1em}$<9$th Grade & $1.32$ ($-3.33$, $5.97$) & $1.06$ ($-3.64$, $5.77$) & $1.18$ ($-3.48$, $5.84$) & $1.24$ ($-3.42$, $5.91$)\\
\hspace{1em}$9-11$th Grade & $2.97$ ($-0.72$, $6.65$) & $3.05$ ($-0.68$, $6.77$) & $3.04$ ($-0.65$, $6.73$) & $3.16$ ($-0.52$, $6.85$)\\
\hspace{1em}High School Grad/GED or Equivalent & $3.07$ ($0.76$, $5.37$) & $2.74$ ($0.36$, $5.11$) & $2.98$ ($0.66$, $5.3$) & $2.76$ ($0.43$, $5.09$)\\
\hspace{1em}Some College or AA Degree & $2.21$ ($0.18$, $4.24$) & $2.12$ ($0.07$, $4.18$) & $2.25$ ($0.21$, $4.29$) & $2.18$ ($0.14$, $4.23$)\\
\addlinespace
\multicolumn{5}{c}{\textit{Model $5$: Folate ($Y_5$) and Food Folate ($X_5$)}}\\
\addlinespace
Intercept & $373.42$ ($331.41$, $415.43$) & $336.95$ ($287.79$, $386.12$) & $392.27$ ($343.64$, $440.9$) & $376.47$ ($327.51$, $425.44$)\\
Exposure of Interest: Food Folate & $0.01$ ($-0.05$, $0.07$) & $0.13$ ($0.03$, $0.23$) & $-0.06$ ($-0.18$, $0.05$) & $0.00$ ($-0.11$, $0.11$)\\
Female (Referent = Male) & $27.00$ ($7.95$, $46.06$) & $32.72$ ($13.23$, $52.21$) & $24.27$ ($4.95$, $43.59$) & $26.49$ ($7.1$, $45.87$)\\
Age & $3.47$ ($2.89$, $4.04$) & $3.52$ ($2.94$, $4.1$) & $3.45$ ($2.88$, $4.03$) & $3.46$ ($2.88$, $4.04$)\\
\addlinespace[0.3em]
\multicolumn{5}{l}{\textit{Race and Ethnicity (Referent = Non-Hispanic White)}}\\
\hspace{1em}Mexican American & $-48.62$ ($-88.46$, $-8.78$) & $-52.47$ ($-92.66$, $-12.27$) & $-46.64$ ($-86.61$, $-6.67$) & $-48.59$ ($-88.46$, $-8.72$)\\
\hspace{1em}Other Hispanic & $-63.12$ ($-95.73$, $-30.51$) & $-61.38$ ($-94.21$, $-28.55$) & $-63.24$ ($-95.86$, $-30.62$) & $-63.17$ ($-95.79$, $-30.55$)\\
\hspace{1em}Non-Hispanic Black & $-104.87$ ($-137.06$, $-72.67$) & $-101.81$ ($-134.24$, $-69.39$) & $-105.26$ ($-137.51$, $-73.02$) & $-104.99$ ($-137.18$, $-72.79$)\\
\hspace{1em}Other Race & $-56.36$ ($-88.69$, $-24.04$) & $-54.36$ ($-86.8$, $-21.93$) & $-54.14$ ($-86.66$, $-21.62$) & $-56.01$ ($-88.57$, $-23.46$)\\
\addlinespace[0.3em]
\multicolumn{5}{l}{Education Level (Referent = College Graduate or Above)}\\
\hspace{1em}$<9$th Grade & $-18.40$ ($-70.66$, $33.86$) & $-5.99$ ($-59.4$, $47.42$) & $-20.61$ ($-72.93$, $31.72$) & $-18.85$ ($-71.1$, $33.39$)\\
\hspace{1em}$9-11$th Grade & $-44.42$ ($-85.79$, $-3.04$) & $-39.95$ ($-82.07$, $2.17$) & $-45.73$ ($-87.1$, $-4.36$) & $-44.95$ ($-86.41$, $-3.49$)\\
\hspace{1em}High School Grad/GED or Equivalent & $-17.21$ ($-43.15$, $8.72$) & $-11.36$ ($-37.83$, $15.10$) & $-20.29$ ($-46.58$, $6.00$) & $-17.67$ ($-43.84$, $8.49$)\\
\hspace{1em}Some College or AA Degree & $6.64$ ($-16.26$, $29.54$) & $15.50$ ($-8.52$, $39.53$) & $3.49$ ($-19.91$, $26.9$) & $6.20$ ($-16.94$, $29.33$)\\
\bottomrule
\end{tabular}
}
\caption{All fitted models' estimates from the application to error-prone dietary intake exposures in the National Health and Nutrition Examination Survey (NHANES) data. For the gold standard (GS) analysis, all $N = 2388$ individuals were assumed to have validated exposures $\bX$ (simulated and assumed to be obtainable only via validation). Otherwise, a subset of $n = 250$ individuals were assumed to have $\bX$; for all others, the validated exposures $\bX$ were multiply imputed from $\bX^*$ and $\bZ$. Three different validation study designs were considered: simple random sampling (SRS), extreme tail sampling on the primary error-prone exposure $X_1^*$ (ETS-$X_1^*$), and extreme tail sampling on the first principal component of all the error-prone exposures $X_1^*,\dots,X_5^*$ (ETS-$PC_1^*$).}
\end{table} 

\end{document}